%% file: main.tex
 \documentclass[twocolumn]{aastex631}
\usepackage{graphicx}	
\usepackage{amsmath}	

\input{macros}

\begin{document}

\title{Unveiling the Cosmic Chemistry II: ``direct" 
$T_e$--based metallicity of galaxies at 3 $< z <$ 10 with JWST/NIRSpec}

\author[0000-0002-4469-2518]{Priyanka Chakraborty}
\affiliation{Center for Astrophysics $\vert$ Harvard \& Smithsonian, 60 Garden st., Cambridge, MA 02138}
\email{priyanka.chakraborty@cfa.harvard.edu}

\author[0000-0002-5222-1337]{Arnab Sarkar}
\affiliation{Kavli Institute for Astrophysics and Space Research,
Massachusetts Institute of Technology, 70 Vassar St, Cambridge, MA 02139}

\author[0000-0003-4284-4167]{Randall Smith}
\affiliation{Center for Astrophysics $\vert$ Harvard \& Smithsonian, 60 Garden st., Cambridge, MA 02138}

\author[0000-0003-4503-6333]{Gary J. Ferland}
\affiliation{University of Kentucky, Lexington, KY}

\author{Michael McDonald}
\affiliation{Kavli Institute for Astrophysics and Space Research,
Massachusetts Institute of Technology, 70 Vassar St, Cambridge, MA 02139}

\author{William Forman}
\affiliation{Center for Astrophysics $\vert$ Harvard \& Smithsonian, 60 Garden st., Cambridge, MA 02138}

\author[0000-0001-8593-7692]{Mark Vogelsberger}
\affiliation{Kavli Institute for Astrophysics and Space Research,
Massachusetts Institute of Technology, 70 Vassar St, Cambridge, MA 02139}

\author[0000-0002-5653-0786]{Paul Torrey}
\affiliation{University of Virginia, Virginia, USA}

\author[0000-0002-8111-9884]{Alex M. Garcia}
\affiliation{University of Virginia, Virginia, USA}

\author[0000-0002-1379-4482]{Mark Bautz}
\affiliation{Kavli Institute for Astrophysics and Space Research,
Massachusetts Institute of Technology, 70 Vassar St, Cambridge, MA 02139}

\author[0000-0003-3462-8886]{Adam Foster}
\affiliation{Center for Astrophysics $\vert$ Harvard \& Smithsonian, 60 Garden st., Cambridge, MA 02138}

\author[0000-0002-3031-2326]{Eric Miller}
\affiliation{Kavli Institute for Astrophysics and Space Research,
Massachusetts Institute of Technology, 70 Vassar St, Cambridge, MA 02139}

\author[0000-0002-4737-1373]{Catherine Grant}
\affiliation{Kavli Institute for Astrophysics and Space Research,
Massachusetts Institute of Technology, 70 Vassar St, Cambridge, MA 02139}

\begin{abstract}

We report the detection of the 
[$\oiii$] auroral line in 42 
galaxies within the redshift 
range of $3 < z < 10$. 
These galaxies were selected
from publicly available JWST 
data releases, 
including the JADES and PRIMAL
surveys, and
observed using both 
the low-resolution 
PRISM/CLEAR configuration 
and medium-resolution gratings. 
The measured electron temperatures
in the high-ionization regions
of these galaxies 
 range from 
$T_e$([$\oiii$]) = 12,000 to 24,000 K, 
consistent with temperatures 
observed in 
local metal-poor galaxies and
previous JWST studies. 
In 10 galaxies,
we also detect the
[$\oii$] auroral line, 
allowing us to determine
electron temperatures in the
low-ionization regions,
which range between $T_e$([$\oii$]) = 10,830 and
20,000 K.
The direct-$T_e$-based metallicities 
of our sample span from  
12 + log(O/H) = 7.2 to 8.4, 
indicating these 
high-redshift galaxies are
relatively metal-poor. 
By combining our sample with 
25 galaxies from the literature,
we expand the dataset to a 
total of 67 galaxies within
$3 < z < 10$,
effectively 
more than doubling the previous sample size for
direct-$T_e$ based
metallicity studies. This larger
dataset allow us to derive
 empirical metallicity 
calibration relations 
based exclusively 
on high-redshift galaxies, 
using six key line ratios: 
R3, R2, R23, Ne3O2,
O32, and O3N2.
Notably, we derive a novel 
metallicity calibration relation for the 
first time using high-redshift $T_e$-based metallicities: 
$\hat{R}$ = 0.18log $R2$ + 0.98log $R3$. 
This new calibration significantly
reduces the scatter in high-redshift 
galaxies compared to the $\hat{R}$ 
relation previously calibrated for 
low-redshift galaxies.

\end{abstract}

\keywords{High redshift Galaxies--Chemical abundances}



\section{Introduction}\label{sec:intro}

Determining the gas-phase metallicities 
in galaxies is essential,
as metallicity serves as a sensitive
tracer of the physical mechanisms that
regulate the baryon cycle.
It reflects the complex interplay 
between gas inflows and outflows, 
star formation, and the subsequent 
enrichment of the interstellar medium 
\citep[e.g.,][]{2008MNRAS.385.2181F, 2011MNRAS.415...11D, 2012ceg..book.....M, 2013ApJ...772..119L,2019ARA&A..57..511K}.

The mass-metallicity relation (MZR) was
first identified by 
\citet{1979A&A....80..155L}, 
with subsequent studies using optical 
luminosity as a mass proxy confirming 
a correlation between blue luminosity 
and metallicity in galaxies
\citep[e.g.,][]{1987ApJ...317...82G, 1994ApJ...420...87Z}. 
Advances in stellar population synthesis 
models \citep{2003MNRAS.344.1000B} 
enabled more precise stellar mass measurements, 
leading to the discovery of a robust 
correlation between stellar mass and
gas-phase oxygen abundance in local 
star-forming galaxies \citep{2004ApJ...613..898T}. 
Observations indicate that the MZR evolves 
up to  $z \sim 3.5$, with metallicities 
increasing at lower redshifts for a 
given stellar mass \citep{2008A&A...488..463M}.

The most reliable method for measuring 
gas-phase metallicity is the
“direct–$T_{e}$ method,” which 
determines the electron temperature
($T_{e}$) using collisionally-excited
auroral lines like [$\oiii$]$\lambda4363$ 
\citep{1967ApJ...150..825P, 2002ApJS..142...35K, 2009ApJ...700..309B}. 
The flux ratio of the weak auroral 
line [$\oiii$]$\lambda4363$ to the
stronger nebular line [$\oiii$]$\lambda5007$ 
provides an accurate diagnostic for
$T_{e}$. 
This method is effective because the 
electron temperature of the emitting 
gas is strongly anti-correlated with 
metallicity, 
as metal ions play a significant role 
in radiative cooling. 
Once $T_e$ is determined, 
gas-phase metallicity can be directly
calculated from the flux ratios of 
standard emission lines
\citep[e.g.,][]{2006agna.book.....O,2009ApJ...699L.161Y,2015ApJ...813..126J,2013ApJ...765..140A,2016ApJ...825L..23S,2017PASP..129d3001P,2020MNRAS.491.1427S,2024ApJ...962...24S}.

Strong-line metallicity calibrations
are established by fitting the
relationships between various strong 
optical nebular emission line ratios
and metallicities determined through 
the direct $T_{e}$ method. 
Extensive studies have been conducted 
to calibrate metallicity diagnostics
using strong line ratios in samples of 
$\hii$ regions within nearby 
star-forming galaxies. 
\citep{2004MNRAS.348L..59P, 2006A&A...459...85N, 2013A&A...559A.114M, 2016MNRAS.458.1529B, 2017MNRAS.465.1384C}.
However, studies aimed at deriving 
metallicity calibration relations for
high-redshift star-forming galaxies
($z > 3$) remain sparse.
\citep{2024ApJ...962...24S,2024A&A...681A..70L}.
Some of the widely used strong
line-ratios are 

\begin{equation}\label{eq:r3}
\text{R3} = \frac{[\text{\oiii}]\lambda5007}{H\beta}
\end{equation}

\begin{equation}\label{eq:r2}
\text{R2} = \frac{[\oii]\lambda\lambda3727,3729}{H\beta}
\end{equation}

\begin{equation}\label{eq:r23}
\text{R23} = \frac{[\oii]\lambda\lambda 3727,3729 + [\oiii]\lambda\lambda 4959,5007)}{H\beta}
\end{equation}

\begin{equation}\label{eq:o32}
\text{O32} = \frac{[\text{\oiii}]\lambda5007}{[\oii]\lambda\lambda3727,3729}
\end{equation}

\begin{equation}\label{eq:ne3o2}
\text{Ne3O2} = \frac{\neiii\lambda3870}{[\oii]\lambda\lambda3727,3729}
\end{equation}

\begin{equation}\label{eq:o3n2}
\text{O3N2} = \frac{\oiii\lambda5007/H\beta}{[\nii]\lambda6585/H\alpha}
\end{equation}

The [$\oiii$]$\lambda4363$ line is a faint
emission line, which restricted the use
of the $T_e$ method for measuring metallicity
in galaxies at $z > 3$ before
the James Webb Space Telescope (JWST) era. 
Consequently, this method was primarily
applied to low-redshift,
metal-poor galaxies 
\citep[e.g.,][]{2006A&A...448..955I, 2019A&A...623A..40I} 
or to stacked samples of thousands 
of high-metallicity galaxies 
\citep[e.g.,][]{2017MNRAS.465.1384C}. 
Auroral line measurements were scarce for
galaxies at $z > 1$ \citep{2009ApJ...699L.161Y, 2012MNRAS.427.1953C, 2012MNRAS.427.1973C, 2017PASJ...69...44K, 2018ApJ...859..175B, 2019ApJ...887..168G} 
and even rarer beyond $z > 3$ \citep{2016ApJ...825L..23S,2024ApJ...962...24S}.

The advent of JWST has improved this
situation by enabling more frequent 
detections of auroral lines at $z > 3$ 
\citep{2022A&A...665L...4S, 2023ApJS..269...33N, 2024ApJ...971...43M, 2024ApJ...962...24S, 2024arXiv241206517S}. 
Nonetheless, auroral line detections at 
$z > 3$ with JWST are still limited to 
around 25 galaxies,
posing challenges for establishing
robust high-redshift metallicity calibrations. 
\citet{2024ApJ...962...24S} combined a 
sample of 25 galaxies at $z>2$ observed
with JWST and 21 galaxies within
$1.4 < z < 3.7$ observed with ground-based 
spectroscopy to present the first high-redshift 
metallicity calibrator within $1.4 < z < 8.7$. 
Similarly, \citet{2024A&A...681A..70L} 
used a sample from JADES, CEERS, 
and ERO, also comprising 25 galaxies 
within $3 < z < 10$, for metallicity 
calibration.

In this paper, we significantly 
expand the high-redshift galaxy 
sample by reporting the detection 
of novel [$\oiii$]$\lambda4363$ auroral 
emission lines in 42
galaxies 
within $3 < z < 10$, observed 
using JWST/NIRSpec medium-resolution 
gratings from the JADES and PRIMAL surveys. 
All of these detections are new. 
We use the dust-corrected [$\oiii$]$\lambda4363$ 
to [$\oiii$]$\lambda5007$ line-flux ratio
to derive robust gas-phase oxygen
abundances using the direct method. 
This new sample is combined with 
25 galaxies in the same redshift 
range from previous JWST studies 
with [$\oiii$] $\lambda4363$ detections 
\citep[e.g.,][]{2024ApJ...962...24S, 2023ApJS..269...33N},
bringing the total to 67 galaxies 
within $3 < z < 10$. 
We use this expanded dataset to 
derive empirical high-redshift
metallicity calibrators, 
enabling robust measurement of
gas-phase oxygen abundance from 
rest-frame optical strong 
emission-line ratios,
as described in Equation
\ref{eq:r3}, \ref{eq:r2},
\ref{eq:r23}, \ref{eq:o32},
\ref{eq:ne3o2}, and \ref{eq:o3n2}. 
These metallicity calibration 
relations will facilitate future
studies in measuring gas-phase 
metallicity across a wide
redshift range,
$3 < z < 10$, and over a 
metallicity range of 
$7.2 < 12+\log(\text{O/H}) < 8.4$. 
Our sample nearly triples the previously 
available sample size, providing a stronger 
foundation for metallicity diagnostics 
in high-redshift galaxies.

The paper is organized as follows. 
Section 2 details the observed
data reduction, analysis, 
and spectral fitting techniques. 
In Section 3, we describe the electron 
temperature and density measurements, 
followed by the determination
of “direct” metallicities. 
Section 4 presents the high-redshift 
metallicity calibration relations.
Section 5 discusses the implications
of these metallicity measurements and 
introduces a novel metallicity
calibration relation. 
Finally, Section 6 summarizes our findings.

Throughout this paper, 
we adopt the solar abundance table from 
\citet{2021A&A...653A.141A}, 
the AB magnitude system 
\citep{1983ApJ...266..713O}, 
and cosmological parameters from \citet{2020A&A...641A...6P}:
the Hubble constant 
$H_0 = 67.4 \, \text{km s}^{-1} \text{Mpc}^{-1} $, 
matter density parameter 
$\Omega_M = 0.315 $, and 
dark energy density $ \Omega_\Lambda = 0.685$.

\begin{figure*}
\centering
\begin{tabular}{cc}
\includegraphics[width=0.5\textwidth]{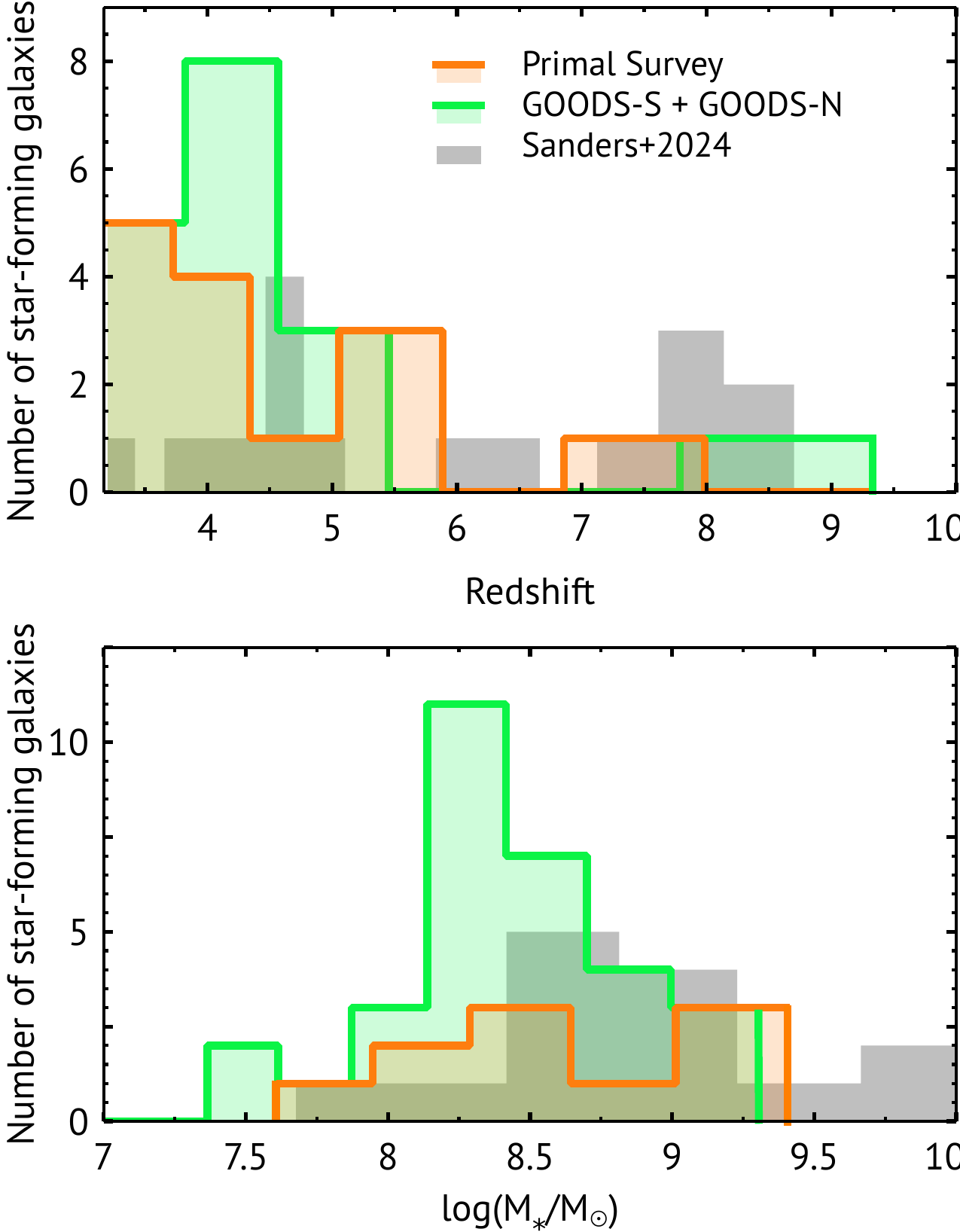} &   \includegraphics[width=0.45\textwidth]{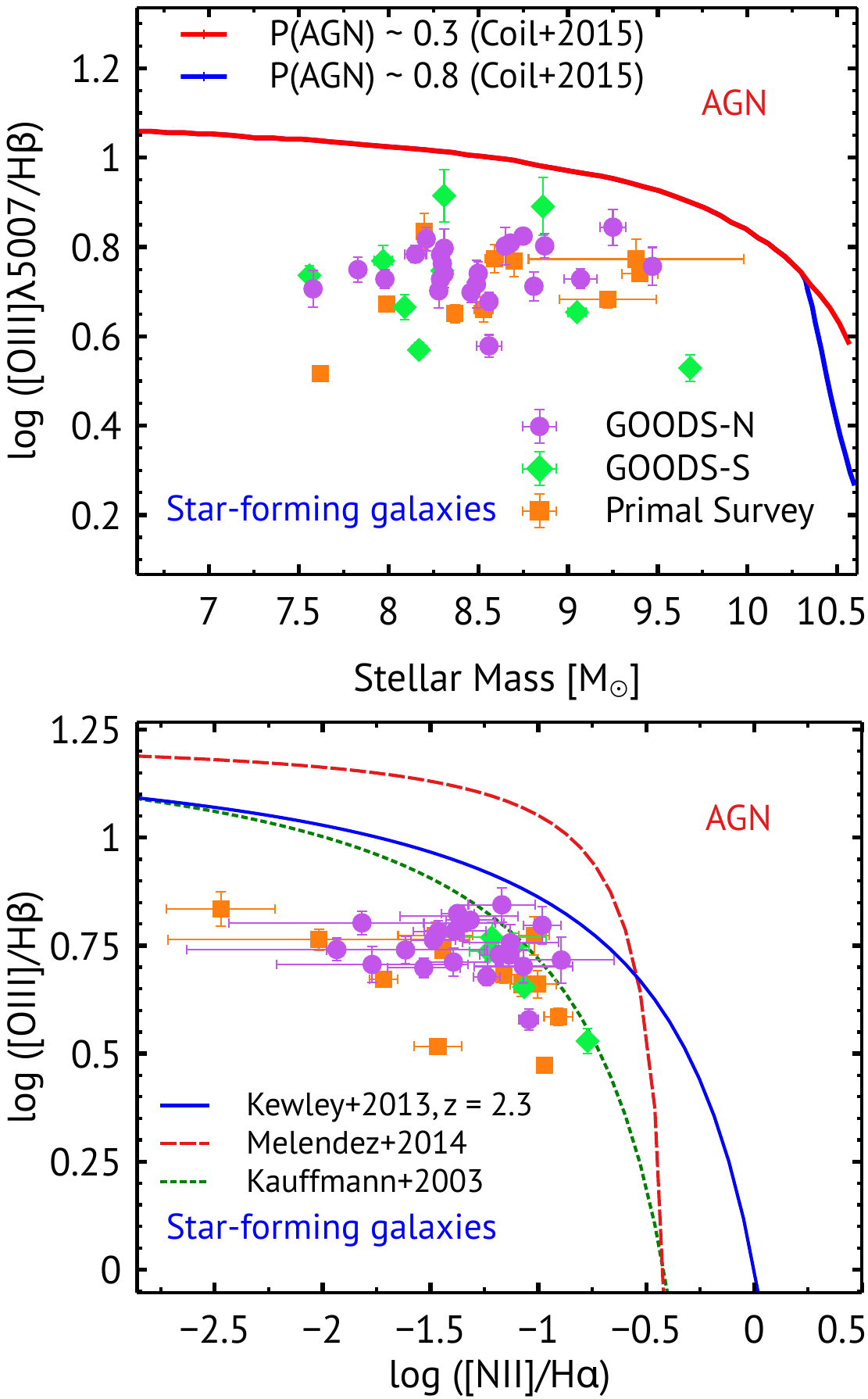}\\
\end{tabular}  
    \caption{Left-top: Redshift distribution ($3 < z < 10$) of galaxies used in our study that exhibit the auroral [$\oiii$]$\lambda4363$ line. The green histogram represents galaxies from GOODS-N and GOODS-S Data Release 3 \citep{2024arXiv240406531D}, while the orange histogram shows galaxies from the PRIMAL survey \citep{2024arXiv240402211H}. The grey histogram captures JWST-detected galaxies from existing literature within $3 < z < 10$ that also display the [$\oiii$]$\lambda4363$ line \citep{2024ApJ...962...24S}.
    { Left-bottom: The stellar mass distribution of galaxies used in our study is shown, using the same color-coding as above.} Right-top: The Mass-Excitation (MEx) diagram for our complete sample, illustrating the relationship between log([$\oiii$]$\lambda$5007/H$\beta$) and the 
    stellar mass.  This diagnostic, based on \citet{2024ApJ...960L..13H} and MEx curves from \citet{2015ApJ...801...35C}, shows probable AGN regions with likelihoods of $\sim$0.3 (red) and $\sim$0.8 (blue). We find no evidence of significant AGN contamination in our sample.
    {Right-bottom: [$\oiii$]/H$\beta$ vs. 
    [$\nii$]/H$\alpha$ BPT diagram for our sample
    of galaxies. The green dotted curve represents the $z\sim0$ demarcation line between star-forming galaxies and AGNs from \citet{2003MNRAS.346.1055K}. The blue solid and red dashed curves indicate the predicted upper limits for star-forming galaxies at $z\sim2.3$, as proposed by \citet{2013ApJ...774L..10K} and the theoretical model from \citet{2014MNRAS.443.1358M}, respectively.}
    }
\label{fig:z_dist_mex}
\end{figure*}

\section{Observations and data analysis}
The data analyzed in this paper were
obtained from multi-object spectroscopy 
observations using the 
micro-shutter assembly (MSA) 
of NIRSpec on JWST 
\citep{2022A&A...661A..81F}. 
Our analysis focuses on spectra
acquired with the NIRSpec
low-resolution
{\it PRISM}/CLEAR configuration, 
which covers a spectral range of
0.6--5.3$\mu$m, as well as with 
medium-resolution 
gratings ($R \sim 1000$), 
specifically G140M/F070LP 
(0.7--1.27$\mu$m), 
G235M/F170LP (1.66--3.07$\mu$m), 
and G395M/F290LP (2.87--5.10$\mu$m).

We used the 
following criteria to
select galaxies for this study:
\begin{itemize}
    \item { Each galaxy must} have been observed with JWST/NIRSpec using the PRISM/CLEAR configuration and, at minimum, medium-resolution gratings that cover the rest-frame wavelength range of 0.35–0.7$\mu$m.

    \item The strong emission lines must be resolved at a $> 3\sigma$ significance level, and the [$\oiii$]$\lambda4363$ auroral line at a $\geq 2\sigma$ significance level.

    \item { Each galaxy must} have a redshift $> 3$, with a detection of the [$\oiii$]$\lambda4363$ auroral line that has not been previously reported.
    
\end{itemize}
We assembled a sample of galaxies 42 galaxies
in the redshift range $3 < z < 10$ that meet the above 
criteria, 
primarily from publicly available 
data releases, 
such as 
32 galaxies from
JADES (DR3; e.g., \citealt{2024arXiv240406531D,2022ARA&A..60..121R,2023arXiv230602465E,2023arXiv230602467B,2023NatAs...7..622C}) and
10 galaxies from
JWST-PRIMAL Legacy Survey 
\citep{2024arXiv240402211H}.
{ Figure \ref{fig:z_dist_mex} (Left-top)
shows the redshift histogram of 
our sample.}

Each target was observed with three
micro-shutters activated,
utilizing a three-point nodding 
sequence along the slit to ensure
comprehensive 
coverage and enhanced data quality. 
For each JADES GOODS-S and GOODS-N target,
flux-calibrated 1D and 2D spectra
were produced by the JADES team
using a custom pipeline developed
by the ESA NIRSpec Science
Operations Team (SOT) and 
Guaranteed Time Observations 
(GTO) teams. 
For detailed descriptions of 
the data reduction steps,
we refer readers to 
\citet{2023arXiv230602467B}, 
\citet{2024A&A...684A..75C}, and 
\citet{2024arXiv240406531D}.
In this paper, we use the reduced 
and flux-calibrated 
{\it medium}-tier 1D and 2D 
spectra of hundreds of targets, 
publicly released as part 
of JADES Data Release
3\footnote{\url{https://jades-survey.github.io/scientists/data.html}} 
\citep{2024arXiv240406531D}.

For targets from the
JWST-PRIMAL Legacy Survey,
we used data from the
DAWN JWST Archive (DJA),
which includes reduced images, 
photometric catalogs, 
and spectroscopic data for 
publicly available JWST data 
products\footnote{\url{https://dawn-cph.github.io/dja}} \footnote{\url{https://s3.amazonaws.com/msaexp-nirspec/extractions/nirspec_graded_v2.html}}. 
The DJA spectroscopic archive 
(DJA-Spec) includes observations 
from major programs, such as CEERS 
\citep{2022ApJ...940L..55F}, GLASS 
\citep{2022ApJ...935..110T}, JADES 
\citep{2023arXiv230602467B}, and UNCOVER 
\citep{2022arXiv221204026B}. 
For further details on the 
data reduction processes, see 
\citet{2024arXiv240402211H}.
Table \ref{tab:goods}
and \ref{tab:primal} lists
all the high-redshift galaxies
studied in this paper.

Additionally, we included 
25 JWST-selected 
star-forming galaxies
from the literature, 
spanning a redshift range of
$3 < z < 10$ and a 
stellar mass range similar to 
that of our sample.
These galaxies display 
[\(\mathrm{O\,III}\)]\(\lambda\)4363 
auroral lines and comprise 16 
galaxies from 
\citet{2024ApJ...962...24S} 
and 9 galaxies from 
\citet{2024A&A...681A..70L}.

\subsection{PRISM/CLEAR spectra}\label{sec:prism_spectra_fitting}
To determine the stellar masses 
of the 42 newly detected 
auroral-line galaxies, 
we fit their PRISM spectra 
using the SED fitting code 
$\texttt{Bagpipes}$ 
\citep{2018MNRAS.480.4379C}. 
$\texttt{Bagpipes}$ generates
detailed model galaxy spectra, 
fitting them to both photometric
and spectroscopic data 
\citep{2008MNRAS.384..449F}, 
and provides posterior 
distributions of galaxy 
properties for each 
source in the sample.
This versatile code can model 
galaxies with various star 
formation histories (SFHs), 
including delayed-$\tau$, constant,
and burst scenarios 
\citep[e.g.,][]{2020ApJ...904...33L, 2024arXiv240605306C}.

In this study, we used a constant 
star-formation model (similar 
to Paper I; 
\citealt{2024arXiv240807974S}), 
allowing star formation ages to 
vary between 0 and 2 Gyr. 
We adopted stellar population 
synthesis models based on the 
2016 version of the BC03 models 
\citep{2003MNRAS.344.1000B}.
These models assume an initial 
mass function (IMF) from 
\citet{2002Sci...295...82K} 
and incorporate nebular line 
and continuum emissions using 
$\Cloudy$ 
\citep{2017RMxAA..53..385F,2023RMxAA..59..327C}.
The SED fitting was conducted 
over a broad parameter range,
with stellar masses
log$(M_{\ast}/M_{\odot})$ 
varying between 4 and 13 and 
stellar metallicities 
log$(Z/Z_{\odot})$ 
ranging from 0.005 to 2.5.
$\texttt{Bagpipes}$ assumes 
solar abundances from 
\citet{1989GeCoA..53..197A} 
and incorporates ISM 
depletion factors, along 
with He and N scaling 
relations from 
\citet{2000ApJ...542..224D}.

The ionization parameter for 
nebular line and continuum 
emissions was varied between 
$-4$ and $-1$. 
We adopted the Calzetti dust
attenuation curve 
\citep{2000ApJ...533..682C}, 
with an extinction parameter 
$A_{V}$ ranging from 0 to 4. 
To account for birth-cloud dust 
attenuation, 
we introduced a multiplicative 
factor ($1 < \eta < 2$) 
in the dust model,
addressing the increased dust
attenuation typically observed 
around $\hii$ regions, 
which is usually double 
that of the general ISM 
during the galaxy's first 
10 Myr \citep{2023arXiv230602467B}. 
To model this effect, 
we capped the maximum age 
of the birth cloud at 0.01 Gyr 
\citep{2023arXiv230602467B}.
The resulting stellar masses 
and other galaxy properties 
are provided in Table 
\ref{tab:goods} and 
\ref{tab:primal}.
{ Figure \ref{fig:z_dist_mex}
(left-bottom) shows the stellar
mass histogram of our sample
of galaxies.}

\subsection{Emission line-flux measurements}
We used medium-resolution grating 
spectra to measure the line fluxes
of several prominent 
nebular emission lines, 
when detected, 
including hydrogen Balmer lines, 
[$\oii$]$\lambda\lambda$3727,3729, 
[$\neiii$]$\lambda$3867, 
[$\oiii$]$\lambda$4363, 
[$\oiii$]$\lambda\lambda$4959,5007, 
and [$\nii$]$\lambda$6584. 
The continuum and line emissions
were modeled simultaneously 
using the {\tt MSAEXP}
python 
modules\footnote{\url{https://github.com/gbrammer/msaexp}}. 
{\tt MSAEXP} models the continuum
as a linear combination of 
simple stellar-population spectra, 
matching the spectral resolution 
of the observed spectrum.
For continuum modeling, 
we used SFHZ 
templates\footnote{\url{https://github.com/gbrammer/eazy-photoz/tree/master/templates/sfhz}} 
derived from the 
Flexible Stellar Population 
Synthesis (FSPS) models from
{\tt EAZY} \citep{2008ApJ...686.1503B}. 
Specifically, we adopted the 
{CORR\_SFHZ\_13} subset of models, 
which feature 
redshift-dependent SFHs that, 
at a given redshift, 
exclude SFHs that begin earlier 
than the age of the universe.
The maximum allowed attenuation 
is also constrained by each epoch.
Additionally, we included the
best-fit template for the 
JWST-observed extreme 
emission-line galaxy 
at $z = 8.5$ (ID4590) from 
\citet{2023Natur.619..716C}, 
rescaled to match the
normalization of the FSPS models,
to account for potential
emission lines with 
large equivalent widths. 
Emission lines were modeled 
using Gaussian profiles centered
at each line’s wavelength.

{ Next, we derived the 
dust-corrected
flux of the key emission lines
implementing 
\citep[e.g.,][]{2023ApJS..269...33N,2024arXiv240807974S}
\begin{equation}
L_{\text{int}}(\lambda) = L_{\text{obs}}(\lambda) 10^{0.4 k(\lambda) E(B-V)},
\end{equation}
where, $L_{\rm int}(\lambda)$ 
and
$L_{\rm obs}(\lambda)$
represent
the intrinsic and 
observed fluxes, respectively,
$k_{\lambda}$ denotes the 
extinction coefficient at wavelength 
$\lambda$, and the 
specific reddening curve $k_{\lambda}$ 
was adopted from 
\citet{2000ApJ...533..682C}.}
We employed three different 
approaches to determine 
the dust-corrected flux:
\begin{itemize}
    \item For galaxies at $z < 6.75$ 
    where both H$\alpha$ and 
    H$\beta$ were detected with
    a signal-to-noise 
    ratio (S/N) $\geq$ 3, 
    we estimated $E(B - V)$ 
    using the Balmer 
    decrement method. 
    We assumed an intrinsic flux ratio of 
    H$\alpha$/H$\beta = 2.86$ 
    \citep{2006agna.book.....O} 
    and applied the dust extinction 
    curve from 
    \citet{2000ApJ...533..682C}.
    
    \item For galaxies at $z \geq 6.75$, 
    where H$\alpha$ is not observable
    due to the spectral coverage 
    of NIRSpec, 
    we instead used the 
    H$\gamma$/H$\beta$ ratio, 
    assuming an intrinsic value
    of H$\gamma$/H$\beta = 0.47$, 
    which corresponds to a 
    temperature of \(10^4\) K 
    for Case B recombination 
    \citep{2006agna.book.....O}.
    
    \item If neither H$\gamma$ nor
    H$\alpha$ was detected, 
    we estimated the nebular dust
    attenuation using SED fitting
    on the {\it PRISM} spectra, 
    performed with the 
    $\texttt{Bagpipes}$ code. 
    This code incorporates a 
    two-component dust attenuation
    model that accounts for
    both nebular and stellar
    emission 
    (see Section 
\ref{sec:prism_spectra_fitting}).
\end{itemize}

To assess the quality
of our measurements, 
we calculated the signal-to-noise ratios 
(S/N) for each emission line 
and established a minimum
S/N threshold of $\geq 3\sigma$ for
including an emission line in
our metallicity calculations. 
Specifically, our selection 
criteria required galaxies 
to exhibit detectable 
H$\beta$ and [$\oiii$], 
along with at least one other
emission line -- [$\oii$]$\lambda\lambda$3727,3729,
[$\nii$]$\lambda$6584, or
[$\sii$]$\lambda\lambda$6716,31 -- 
measured at or above the 3$\sigma$
confidence level. 
This rigorous approach enables
robust measurement and validation
of emission-line fluxes across 
our sample, ensuring that 
only reliable data points 
are used to derive the 
gas-phase metallicities of 
high-redshift galaxies.

\subsection{AGN contamination: MEx diagram and BPT diagram}
In this study,
we apply line-flux ratio 
diagnostics specifically 
designed to measure gas-phase 
metallicity in star-forming 
regions of galaxies. 
However, AGN-driven ionization can
introduce inaccuracies in 
these metallicity calibrations
by contaminating the emission 
line fluxes. 
To ensure precise 
metallicity measurements, 
we carefully examine each galaxy
for possible AGN influence. 
We use three methods to systematically
exclude AGN-contaminated 
sources from our sample,
thereby strengthening the
reliability of our results.

{First, we use the BPT diagram to
differentiate star-forming galaxies 
from AGNs, employing the 
[$\oiii$]/H$\beta$ vs.
[$\nii$]/H$\alpha$ ratios for our
target sample, as shown in Figure 
\ref{fig:z_dist_mex} (bottom-right).
To identify potential AGNs, we apply three demarcation lines from the literature.
We adopt the empirical demarcation
line from \citet{2003MNRAS.346.1055K},
which distinguishes local 
star-forming galaxies from AGNs. 
We then compare our sample with 
the upper limit for star-forming
galaxies at $z\sim2.3$ proposed
by \citet{2013ApJ...774L..10K} 
and the theoretical
classification scheme from 
\citet{2014MNRAS.443.1358M}.
As shown in Figure \ref{fig:z_dist_mex},
four galaxies in our sample lie above the 
demarcation line from 
\citet{2003MNRAS.346.1055K}.
However, all galaxies remain
below the more recent demarcation
lines from \citet{2013ApJ...774L..10K}
and \citet{2014MNRAS.443.1358M}.
Therefore, we do not exclude any
galaxies from our sample.
}

Secondly, we re-check our sample
by employing the 
Mass-Excitation (MEx) diagnostic diagrams,
initially introduced by 
\citet{2014ApJ...788...88J} 
and further refined by 
\citet{2015ApJ...801...35C}, 
which use the 
[$\oiii$]~$\lambda5007$/H$\beta$ (R3)
emission-line ratio in 
conjunction with stellar mass 
to distinguish AGNs 
from star-forming galaxies. 
This diagram provides an 
alternative to the commonly 
used BPT diagram 
\citep[e.g.,][]{1981PASP...93....5B,2013ApJ...774L..10K}, 
which compares the 
[$\oiii$]$\lambda5007$/H$\beta$ 
and [$\nii$]$\lambda6584$/H$\alpha$ 
emission line ratios but can be
limited when [$\nii$] or H$\alpha$
lines fall outside the
spectral coverage or are blended. 
As our measurements rely 
exclusively on emission-line
fluxes from medium-resolution
grating spectra, 
we use dust-corrected 
[$\oiii$]$\lambda5007$ and H$\beta$
fluxes for the MEx diagram.
Figure \ref{fig:z_dist_mex} (right)
displays 
our JWST samples in the 
log([$\oiii$]$\lambda5007$/H$\beta$) 
vs. log($M_{\ast}/M_{\odot}$) plane.

In Figure \ref{fig:z_dist_mex} 
(right), 
the blue and red curves mark 
steep gradients at P(AGN)
$\sim$ 0.3 and P(AGN) $\sim$ 0.8, 
respectively, representing 
the probability that a
galaxy hosts an AGN, 
as established by 
\citet{2015ApJ...801...35C} 
for $z = 2.3$ galaxies and 
AGNs in the MOSDEF survey 
\citep{2024ApJ...960L..13H}. 
The positions of our sources 
on the MEx diagram indicate 
that our sample is 
predominantly composed of
star-forming galaxies, 
lying below or near
the boundary line. 
Consequently, no galaxies
were excluded due to potential
AGN contamination, and we 
retain all sources in our sample.
A similar method was also used by 
\citet{2024ApJ...960L..13H}
to distinguish star-forming 
galaxies from AGNs in 
the GLASS-JWST sample. 

Additionally, we visually 
inspected each spectrum for 
signs of broad emission
line regions, 
following procedures from 
\citet{2021ApJ...907...12S}.

\subsection{Samples with $[\oiii]\lambda4363$ auroral line}\label{sec:auroral_line_sample}
\begin{figure*}
    \centering
    \includegraphics[width=1\textwidth]{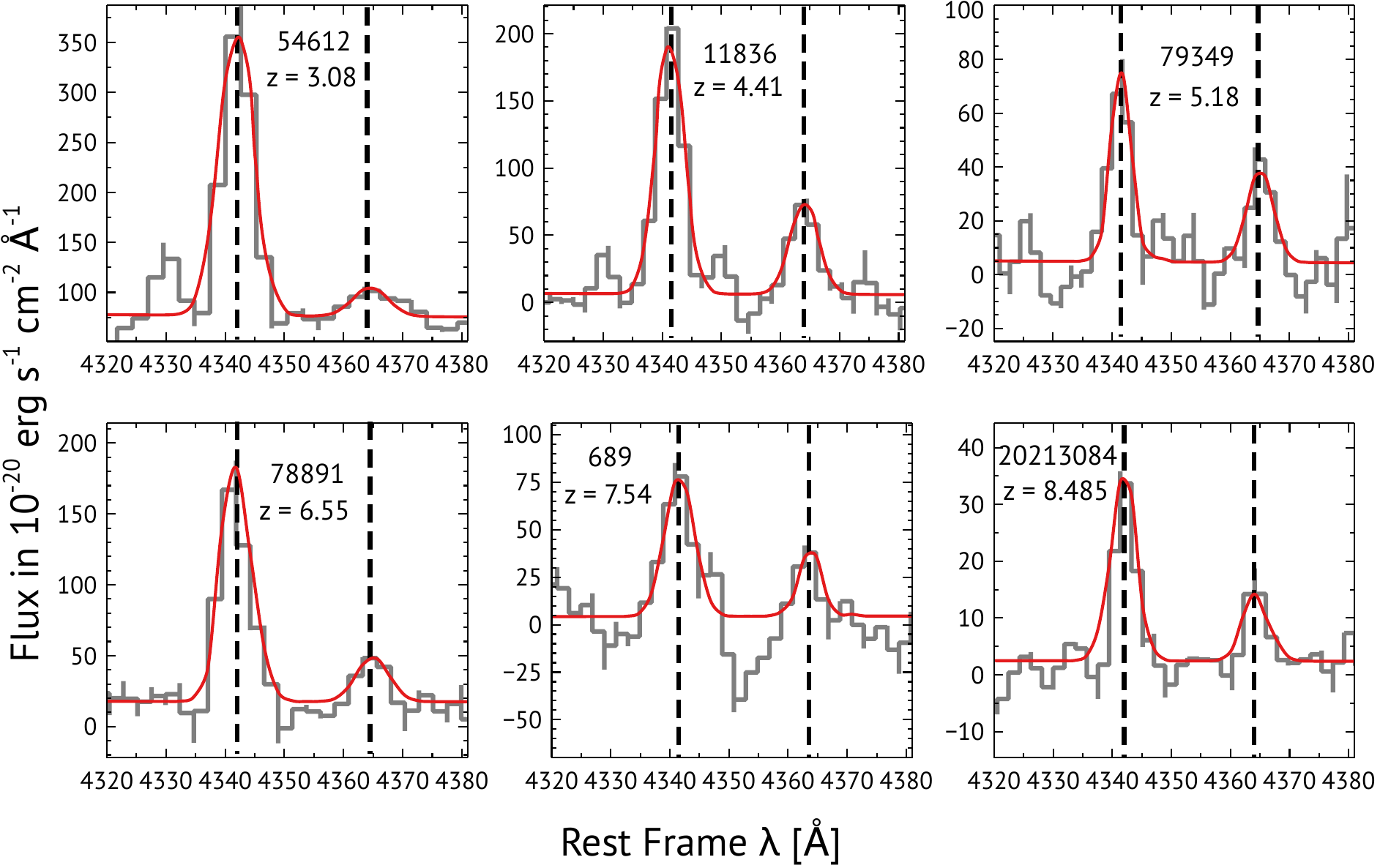}
    \caption{The 1D spectra of six galaxies
    in our sample
    illustrate the detected 
    [$\oiii$]$\lambda$4363 and H$\gamma$ 
    emission lines, 
    representing galaxies
    across various redshifts. 
    {The x-axes are converted to
    rest-frame wavelengths
    using the best-fit redshifts.}
    The red curve shows 
    the best-fit continuum
    along with the 
    [$\oiii$]$\lambda$4363 
    and H$\gamma$ emission
    line profiles. 
    The black dashed line 
    marks the rest-frame 
    wavelengths of
    [$\oiii$]$\lambda$4363
    and H$\gamma$.
    }
\label{fig:spectra_o3_auroral}
\end{figure*}

In this study, we identified 
star-forming galaxies displaying 
an [$\oiii$]$\lambda$4363 auroral 
line with a signal-to-noise
ratio above 2. 
We utilized publicly 
available, reduced,
and calibrated 1D
grating spectra from
the JADES and PRIMAL Surveys
to assemble a sample of 42
galaxies with significant 
[$\oiii$]$\lambda$4363 
emission lines. 
Figure 
\ref{fig:spectra_o3_auroral} 
illustrates a zoomed-in view of 
the observed [$\oiii$]$\lambda$4363 lines 
with the best-fit continuum and 
emission-line models.
Furthermore, we visually inspected
both the 1D and 2D spectra 
for each galaxy to verify
the absence of single-pixel noise
artifacts at the location of the
[$\oiii$]$\lambda$4363 line.

{ In this study, we broadened our
sample by 
combining 
42 newly identified galaxies
with auroral line emissions
with 25 additional galaxies 
from existing literature, }
all exhibiting significant 
[$\oiii$]$\lambda$4363 line fluxes
(S/N $\geq$ 2).
This supplementary group includes
16 galaxies from 
\citet{2024ApJ...962...24S}, 
four galaxies from the 
SMACS 0723 galaxy cluster field 
within the ERO program, 
and five galaxies from the 
GLASS ERS program 
\citep{2022ApJ...935..110T}.

To estimate the $T_{\rm e}$-based 
metallicity for these 25 galaxies,
we used the emission line fluxes 
reported in the respective studies. 
Specifically, for galaxies selected
from the ERO and GLASS ERS programs,
we referred to line fluxes provided in 
\citet{curti2023jadesinsightslowmassend} 
(for ERO IDs 4590, 6355, and 10612), 
\citet{2023ApJS..269...33N} 
(for ERO IDs 5144 and 
GLASS IDs 100003, 10021, 150029,
and 160133), and 
\citet{2023ApJ...951L..17J} 
(for GLASS ID 150008). 
This results in a robust 
sample of 67 star-forming galaxies 
with detected auroral lines, 
representing a sample size that is
at least 2.6 times larger than 
that used in previous direct 
$T_{\rm e}$-based metallicity 
studies of high-redshift 
JWST galaxies.

\section{``direct'' metallcity measurements}\label{sec:metal_measure}
We derived electron temperatures 
and used them to estimate oxygen 
abundances,
with oxygen serving as a proxy for
total gas-phase metallicity. 
This method assumes proportional 
scaling of all elements and models 
each galaxy as a single $\hii$ region, 
divided into a high-ionization zone 
traced by O$^{2+}$ and a 
low-ionization zone traced by O$^{+}$. 
However, this simplified approach
does not account for the 
complexities of temperature 
distributions and ionization 
structures within these regions. 
While a full exploration of these 
assumptions and their limitations 
is beyond the scope of this study,
we recommend 
\citet{2002RMxAC..12...62S},
\citet{2019A&ARv..27....3M},
and 
\citet{2023MNRAS.522L..89C}
for further details on these topics.

\subsection{Electron temperatures; $T_{e}$($\oii$) 
vs $T_{e}$($\oiii$)}
The thermal structures of $\hii$ 
regions can be characterized by 
measuring the fluxes of 
their emission lines. 
The direct method requires the 
detection of at least one auroral
emission line per ion to determine
the temperatures of different ionization zones. 
Ideally, a complete ionization 
structure of an $\hii$ region would
be necessary to measure electron 
temperatures and densities across all zones; 
however, this is impractical, 
so simplified models are commonly 
applied. 
Most studies adopt either a 
two-zone or three-zone model to 
approximate the $\hii$ regions 
that produce the observed 
emission lines in galaxies.
Here, we use a two-zone model 
for $\hii$ regions: (1) an inner
high-ionization zone, characterized by 
$T_e$([$\oiii$]), and (2) an outer 
low-ionization zone, represented
by $T_e$([$\oii$]) 
\citep{1992AJ....103.1330G}.

The $T_e$([$\oiii$]) can be derived 
from the emission line ratio 
of $[\oiii]\lambda5007$ to 
$[\oiii]\lambda4363$ \citep[e.g.,][]{2006agna.book.....O,2017PASP..129d3001P}.
We used {\tt PyNeb} 
\citep{2015A&A...573A..42L,2020Atoms...8...66M} 
to fit the relationship between
this line ratio and
electron temperature, 
assuming a five-level atom
model and adopting collisional
strengths from 
\citet{2014MNRAS.441.3028S}. 
Since $T_e$([$\oiii$]) has only a
weak dependence on electron
density ($n_e$), 
we assume $n_e = 300$ cm$^{-3}$ 
for simplicity in the 
high-ionization zones. 
Testing this with $n_e = 1000$
cm$^{-3}$ yielded only a minor 
temperature reduction of
0.1\% \citep{2020MNRAS.491.1427S}.
For our sample of 42 galaxies,
the derived $T_e$([$\oiii$]) values 
range from approximately
12,000 K to 24,000 K 
(see Table \ref{tab:goods}
and \ref{tab:primal}).
Temperature uncertainties were 
estimated using the affine 
invariant Markov Chain Monte Carlo (MCMC)
sampler in the {\tt emcee} 
package \citep{2013PASP..125..306F}. 
Our $T_e$([$\oiii$]) measurements
are consistent with those of 
extremely metal-poor 
local galaxies, 
as well as with high-redshift
galaxies at $z \sim 6$–8 
\citep[e.g.,][]{2022ApJ...940L..23A, 2022A&A...665L...4S,curti2023jadesinsightslowmassend} 
and galaxies from CEERS at $z \sim 2$–9 
\citep{2024ApJ...962...24S}.

\begin{figure*}
    \centering
\includegraphics[width=1.0\textwidth]{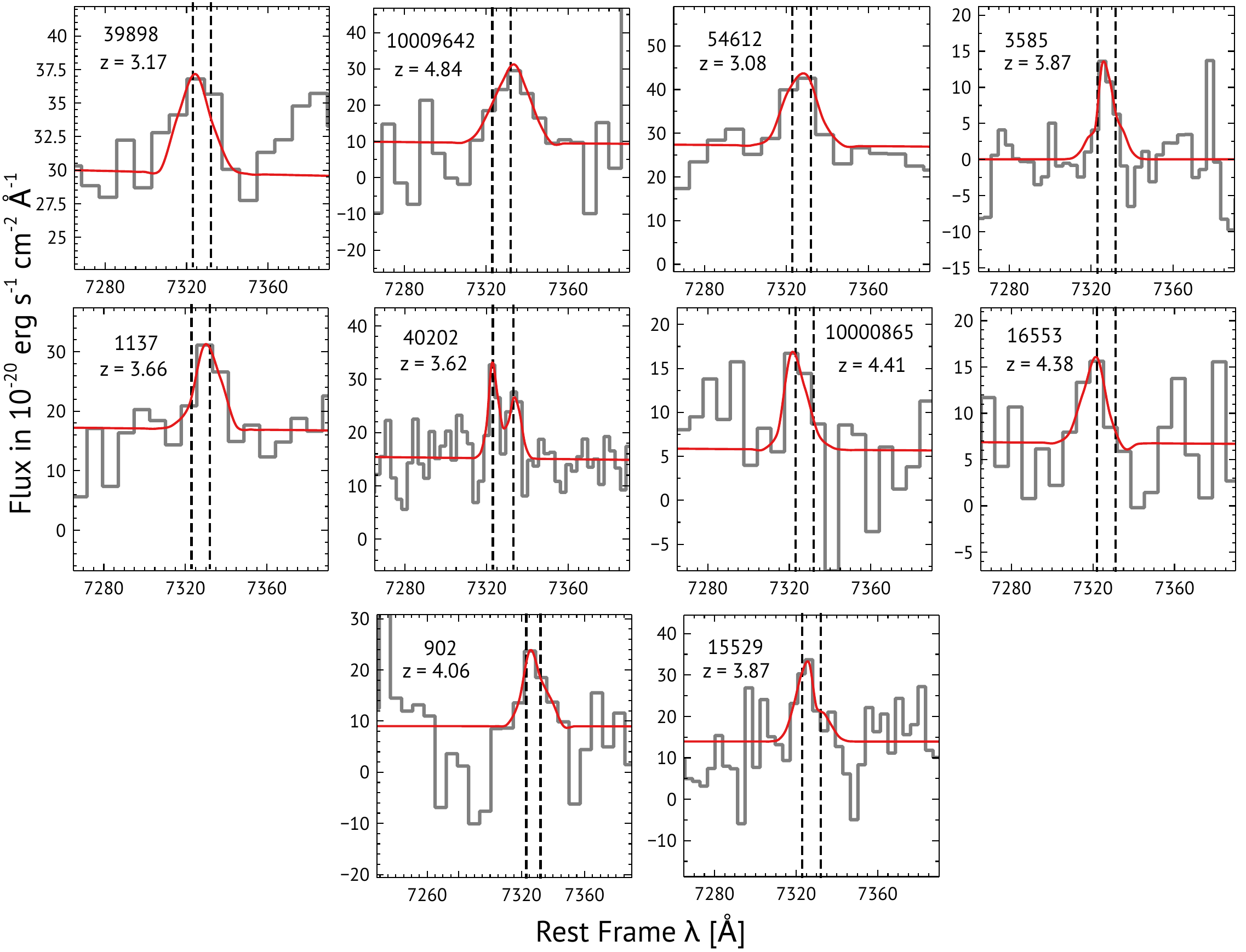}
    \caption{The 1D spectra of 10 
    galaxies in our sample show
    the detected
    [$\oii$]$\lambda\lambda7322, 7332$ 
    doublet. 
     { The x-axes are converted to
    rest-frame wavelengths
    using the best-fit redshifts.}
    The grey represents 
    the observed data, 
    while the red curve displays
    the best-fit continuum along
    with the emission line
    profiles for 
    [$\oii$]$\lambda7322$ and 
    [$\oii$]$\lambda7332$. 
    Vertical black dashed lines
    indicate the rest-frame 
    wavelengths of these lines.
    }
    \label{fig:o2_auroral_lines}
\end{figure*}

The $T_e$([$\oii$]) can be estimated 
from the emission line ratio of
$[\oii]\lambda\lambda 3726, 3729$ 
to $[\oii]\lambda\lambda 7322, 7332$ 
\citep{2006agna.book.....O}. 
This calculation is particularly
challenging due to (1) the 
strong dependence of the $[\oii]$ 
line ratio on electron density, 
and (2) the limited JWST/NIRSpec
spectral coverage of the 
$[\oii]\lambda\lambda 7322, 7332$ 
auroral lines for galaxies at
$z \geq 5.75$. 
Within our sample of 42 galaxies, 
we detect $[\oii]\lambda\lambda 7322, 7332$ 
auroral lines with a 
signal-to-noise ratio (S/N) $\geq 2$ 
in only 10 galaxies.
Figure \ref{fig:o2_auroral_lines} 
presents
the $[\oii]\lambda\lambda 7322, 7332$
emission lines along with
their best-fit
continuum + line-emission
models.

\begin{figure}
    \centering
   \includegraphics[width=0.45\textwidth]{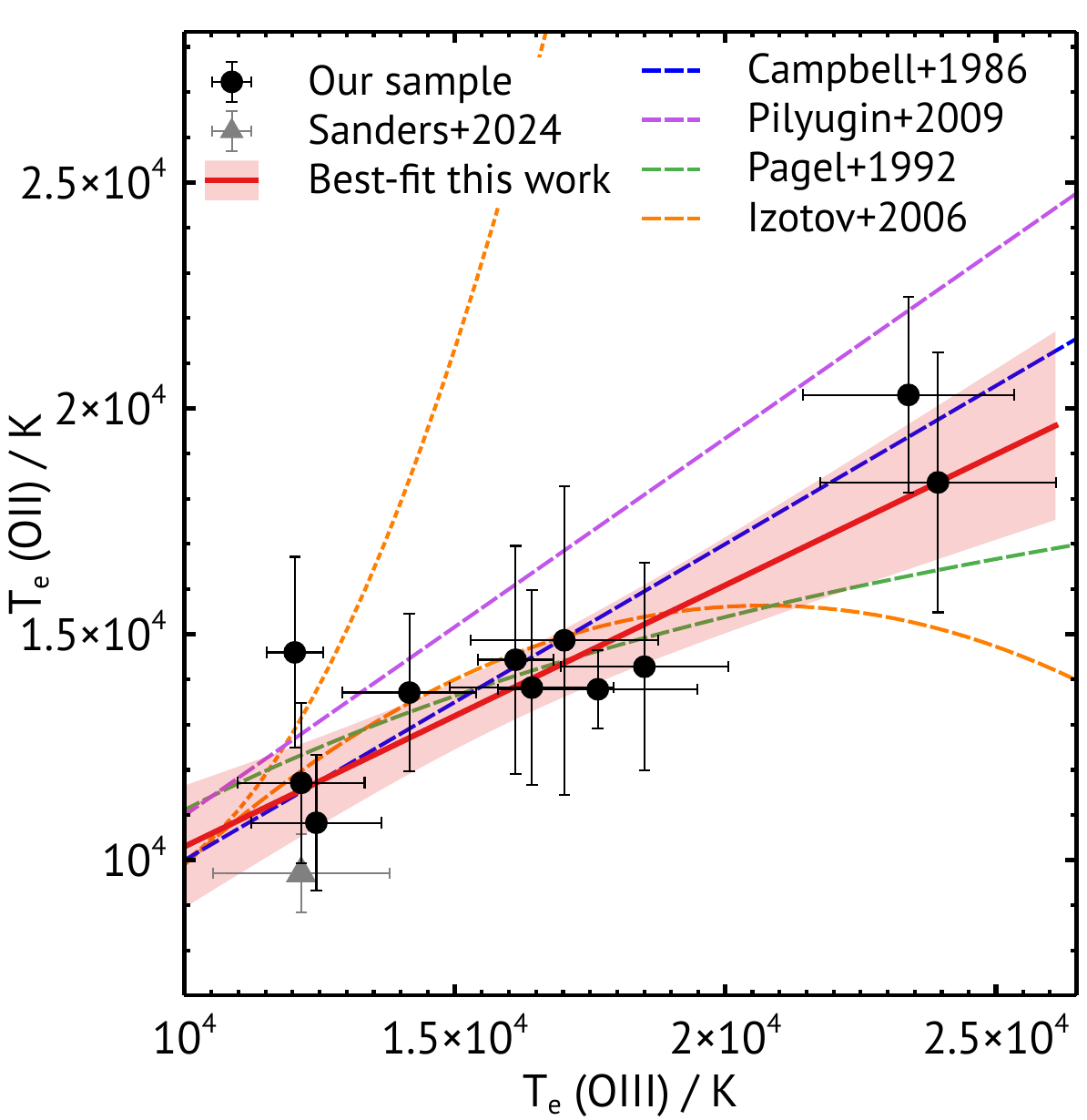}
    \caption{The relation between
    the electron temperatures, 
    $T_e$([$\oii$]) 
    and $T_e$([$\oiii$]), 
    is shown for the 10 galaxies 
    in our sample with detected 
    $[\oii]\lambda\lambda 7322, 7332$ 
    auroral lines. 
    Black filled circles 
    represent our sample,
    while the gray triangle 
    indicates the
    one galaxy from 
    \citet{2024ApJ...962...24S}
    at $z$ = 3.302. 
    The red curve shows
    the best-fit linear 
    regression line with 
    $1\sigma$ uncertainties.
    Blue, violet, and green 
    dashed lines display 
    $T_e$([$\oii$]) vs. $T_e$([$\oiii$]) 
    relations for local galaxies 
    from 
    \citet{1986MNRAS.223..811C}, 
    \citet{2009MNRAS.398..485P}, and 
    \citet{1992MNRAS.255..325P}. 
    The orange dashed and dotted curves, 
    adapted from 
    \citet{2006A&A...448..955I}, 
    represent metallicities $12 + 
    \log(\text{O/H}) = 7.2$ and $8.2$, 
    respectively.
    }
    \label{fig:t2_vs_t3}
\end{figure}

To estimate $T_e$([$\oii$]) in 
the 10 galaxies, we first 
derived $n_e$ from the 
$[\sii]\lambda\lambda 6716, 6731$ 
doublet line ratios
(detected with S/N $\geq 3$) using {\tt PyNeb}, 
assuming an electron temperature 
of $10^4$ K.
{This step is minimally affected by temperature variations, as the resulting $T_e$([$\oii$])
remains largely unchanged even if 
the electron temperature varies by
a factor of 2--3.
}
The measured $n_e$ spans from
approximately 50 cm$^{-3}$ to
1862 cm$^{-3}$ across these galaxies, 
with a median of $567 
\pm 170$ cm$^{-3}$. 
We then derived $T_e$([$\oii$]) from
the $[\oii]\lambda\lambda 3726, 3729$ 
to $[\oii]\lambda\lambda 7322, 7332$ 
line ratios, using the measured
$n_e$ values in {\tt PyNeb} and
adopting O$^+$ collision strengths
from 
\citet{2009MNRAS.397..903K}. 
The resulting $T_e$([$\oii$]) values
range from 10,830 K to 20,000 K
across the subsample. 
Figure \ref{fig:t2_vs_t3} 
displays $T_e$([$\oii$]) as a
function of $T_e$([$\oiii$]) 
for these galaxies.
We find the best-fit relation of:
\begin{equation}\label{eq:o2_best_fit}
    T_e(\oii) = (0.58 \pm 0.19) \times T_e(\oiii) + 
    (4520 \pm 2000) K.
\end{equation}
Our best-fit relation aligns well
with the widely used
$T_e$([$\oii$]) vs. $T_e$([$\oiii$]) 
relation for local galaxies from 
\citet{1986MNRAS.223..811C}. 
It also shows consistency with
the relation by 
\citet{1992MNRAS.255..325P} 
for $T_e$([$\oiii$]) $\leq$ 
23,000 K and with 
\citet{2006A&A...448..955I} 
for low metallicities within 
$T_e$([$\oiii$]) $\leq$ 21,000 K.

As shown in Figure \ref{fig:t2_vs_t3}
and discussed by 
\citet{2024ApJ...962...24S}, 
there is considerable uncertainty 
in the $T_e$([$\oii$]) vs. $T_e$([$\oiii$]) 
relation, even at
low redshift ($z = 0$),
and this relationship displays a
substantial intrinsic scatter.

{To assess the robustness of our metallicity measurements, we have adopted three different relations between $T_e$([$\oii$]) and $T_e$([$\oiii$]): (1) Equation \ref{eq:o2_best_fit}, (2) $T_e$([$\oii$]) = 0.7 $\times$ $T_e$([$\oiii$]) + 3000K (see \citealt{1986MNRAS.223..811C}), (3) $T_e$([$\oii$]) = $T_e$([$\oiii$]).}
These variations did not lead 
to significant changes 
in the metallicity estimates, 
suggesting that the oxygen 
abundance is primarily 
governed by the 
high-ionization region.

\subsection{Oxygen abundances}

We measured ionic oxygen abundances 
using {\tt PyNeb}, 
applying collision strengths
for O$^{2+}$ from 
\citet{2014MNRAS.441.3028S} and 
for O$^+$ from 
\citet{2009MNRAS.397..903K}. 
The electron density was set to 
the median $n_e$ value of 567 
cm$^{-3}$, as determined for 
10 galaxies in our sample. 
This density value aligns
with typical $n_e$ values observed
in galaxies at redshifts $z = 2$–3 
\citep{2016ApJ...825L..23S}.

Since oxygen abundances in the 
interstellar medium (ISM) are 
predominantly influenced by O$^{2+}$ ions,
the effect of variations in 
$n_e$ on abundance measurements is minimal, 
as $T_e$([$\oiii$]) has only a weak
dependence on $n_e$ 
\citep[e.g.,][]{2006A&A...448..955I,2013ApJ...765..140A,2017MNRAS.465.1384C,2020MNRAS.491.1427S}.
Our choice of density is consistent
with studies by 
\citet{2024ApJ...962...24S} 
and \citet{2024A&A...681A..70L}, 
who also fixed $n_e$ 
values
in their $T_e$([$\oiii$])
and metallicity calculations.

We assumed that within the 
$\hii$
regions,
all oxygen exists in either 
the O$^{2+}$ or O$^+$ states, 
allowing us to simplify 
the calculation of the 
total oxygen abundance (O/H).:
\begin{equation}
   \rm \frac{O}{H} = \frac{O^{2+}}{H} + \frac{O^+}{H}.
\end{equation} 
{ Due to the close 
ionization potentials of 
$\heii$ ($\sim$54.4 eV) and 
O$^{3+}$ ($\sim$55 eV), 
we did not apply an ionization
correction factor (ICF) for O$^{3+}$, 
as its inclusion would have only 
a minimal effect on the total 
oxygen abundance 
\citep[e.g.,][]{2006A&A...448..955I,2013ApJ...765..140A,2021ApJ...922..170B,2017MNRAS.465.1384C,2023MNRAS.518..425C}.
\citet{2018ApJ...859..164B} 
found an ICF of 1.055
for unseen O$^{3+}$ ion.
The contribution from O$^{3+}$ 
is thus negligible (0.2--0.5\% in
12+log(O/H))
relative to the uncertainties 
($\sim$ 2--3\%)
in our total oxygen measurements
\citep{2021ApJ...922..170B,2024ApJ...962...24S}. }

The O$^{2+}$/H ratio was calculated 
from the observed
$[\oiii]\lambda5008$/H$\beta$ 
ratio using $T_e$([$\oiii$]), 
while O$^{+}$/H was derived 
from the dust-corrected 
$[\oiii]\lambda\lambda3726, 29$/H$\beta$ 
ratio, 
applying $T_e$([$\oii$]) when 
directly measurable or estimated
via Equation \ref{eq:o2_best_fit} 
otherwise.
Uncertainties were 
assessed using the affine invariant
Markov Chain Monte Carlo (MCMC) sampler
implemented in the {\tt emcee} package
\citep{2013PASP..125..306F},
performing 10,000 realizations based 
on normal distributions of the 
measured line fluxes for
$[\oiii]\lambda\lambda5007, 4959, 4363$, H$\beta$, 
and $[\oiii]\lambda\lambda3726, 29$ with
their associated 1$\sigma$ uncertainties. 
The metallicities for our sample of 
galaxies range from 
$12+\log(\text{O/H}) = 7.2$ to $8.4$, 
suggesting that these high-redshift 
galaxies are relatively metal-poor,
consistent with previous findings 
from JWST observations. 
The metallicities and their 
1$\sigma$ uncertainties are
listed in Table \ref{tab:goods}
and \ref{tab:primal}.

\begin{figure*}
\centering
\begin{tabular}{ccc}
\includegraphics[width=0.33\textwidth]{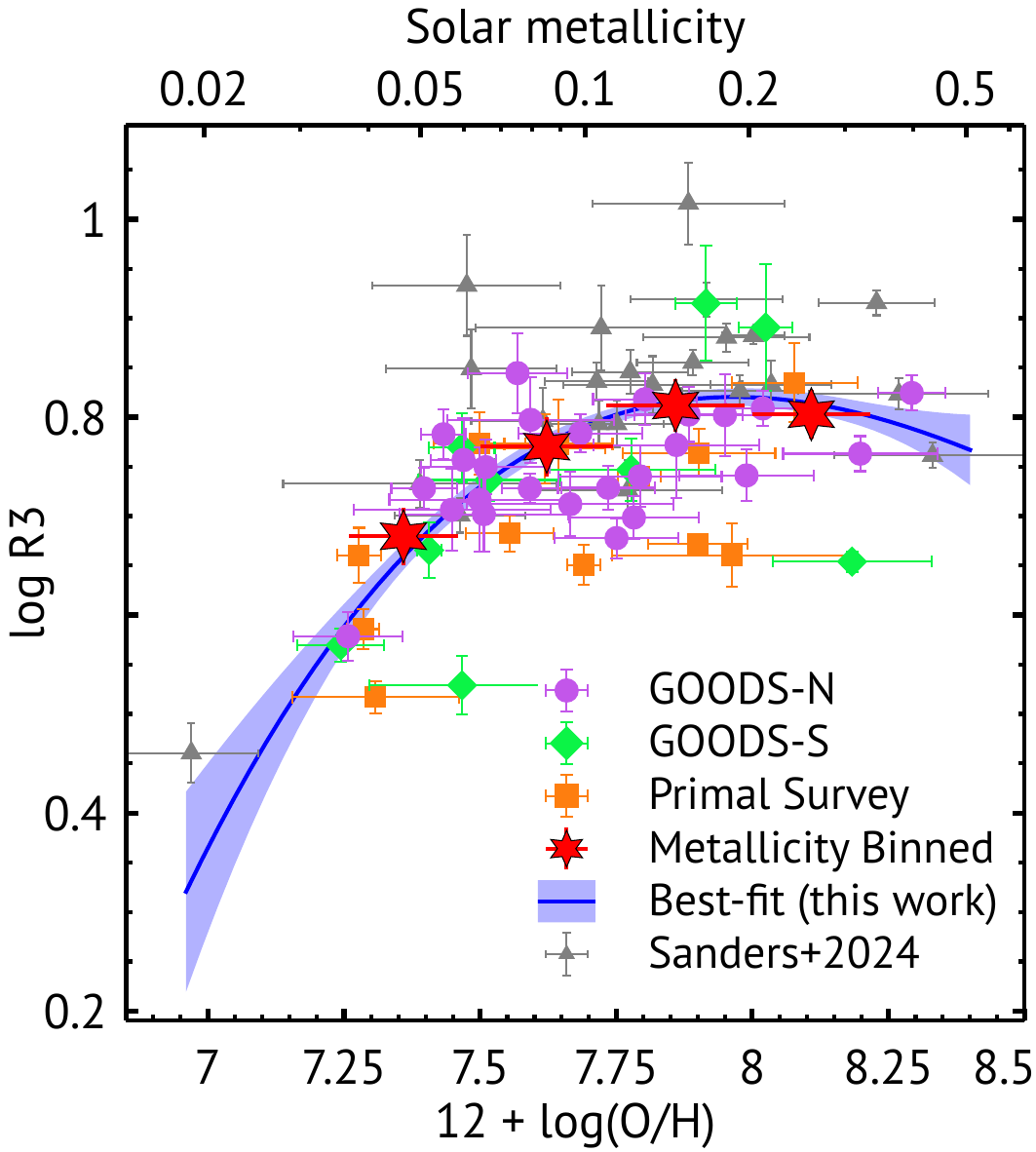} &   \includegraphics[width=0.33\textwidth]{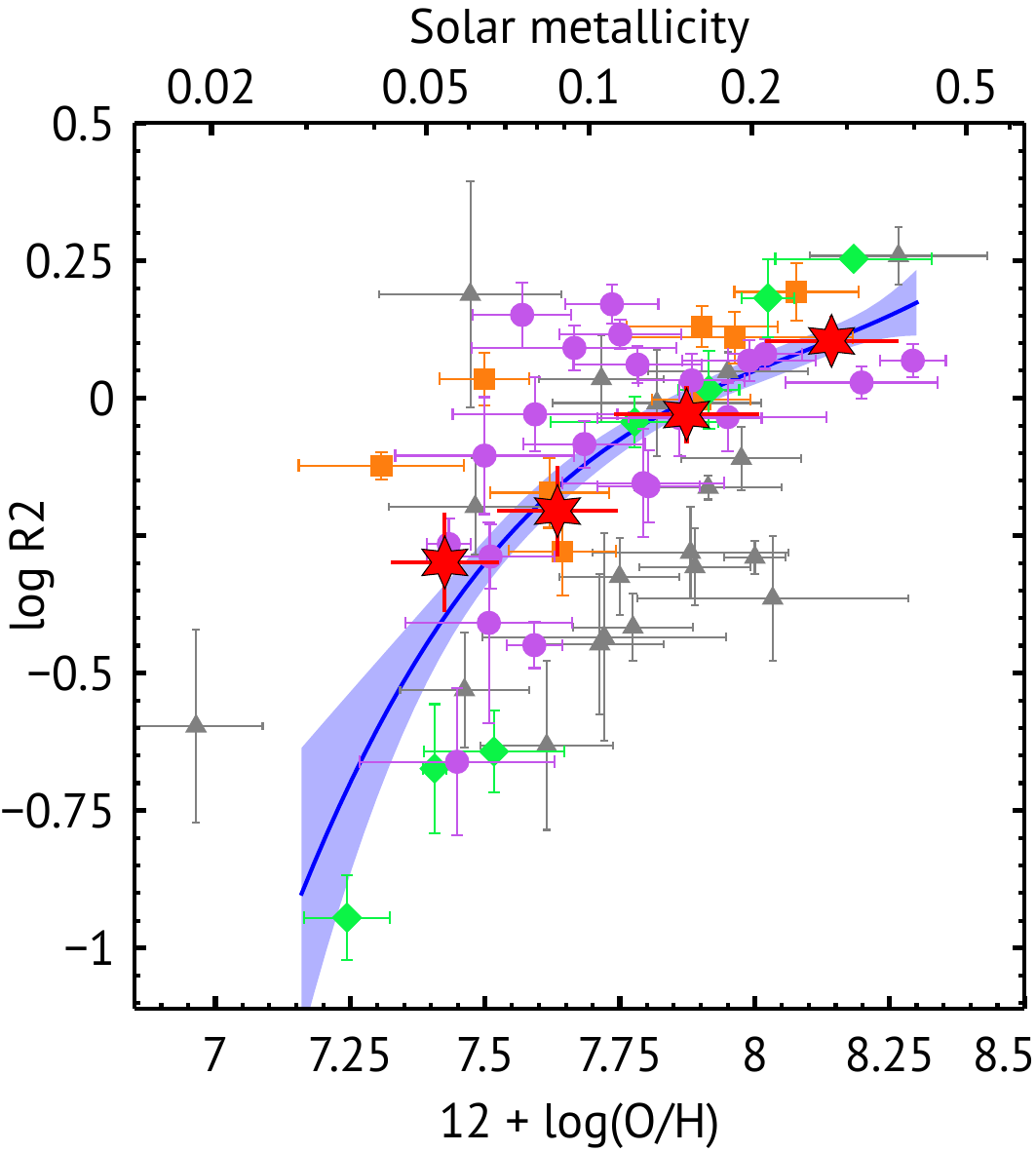}&\includegraphics[width=0.33\textwidth]{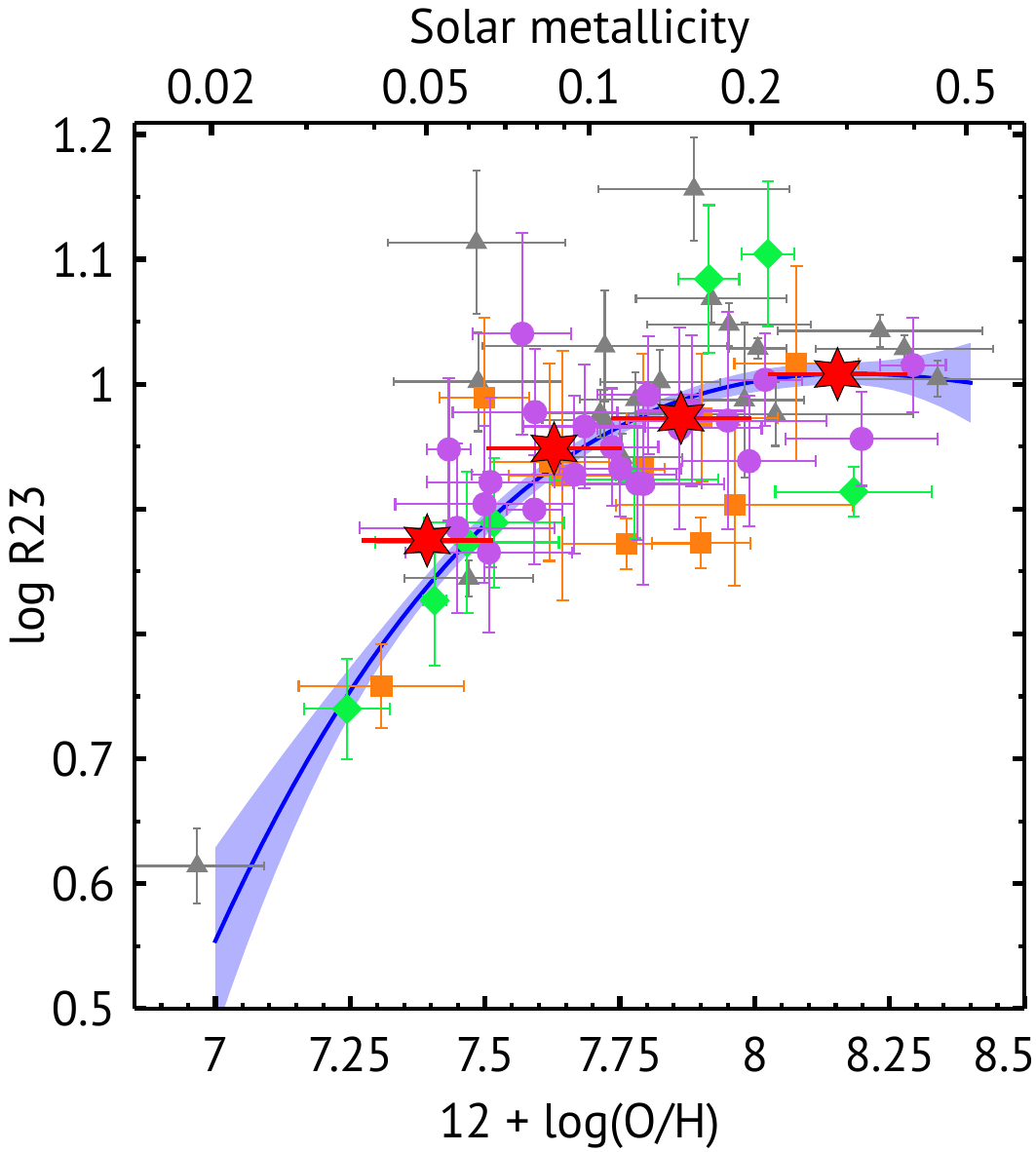}\\
\end{tabular}  
\begin{tabular}{ccc}
\includegraphics[width=0.33\textwidth]{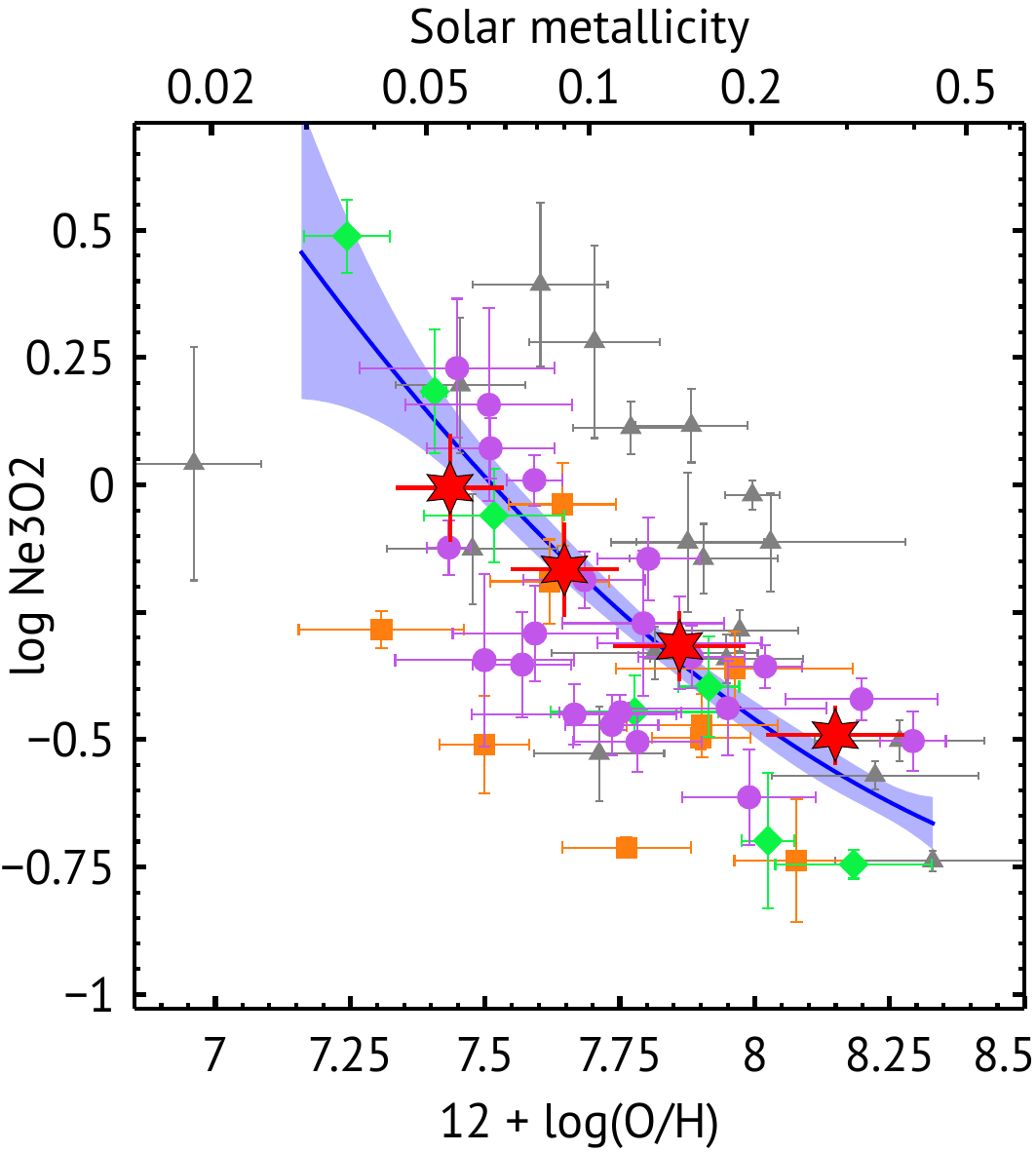} &   \includegraphics[width=0.33\textwidth]{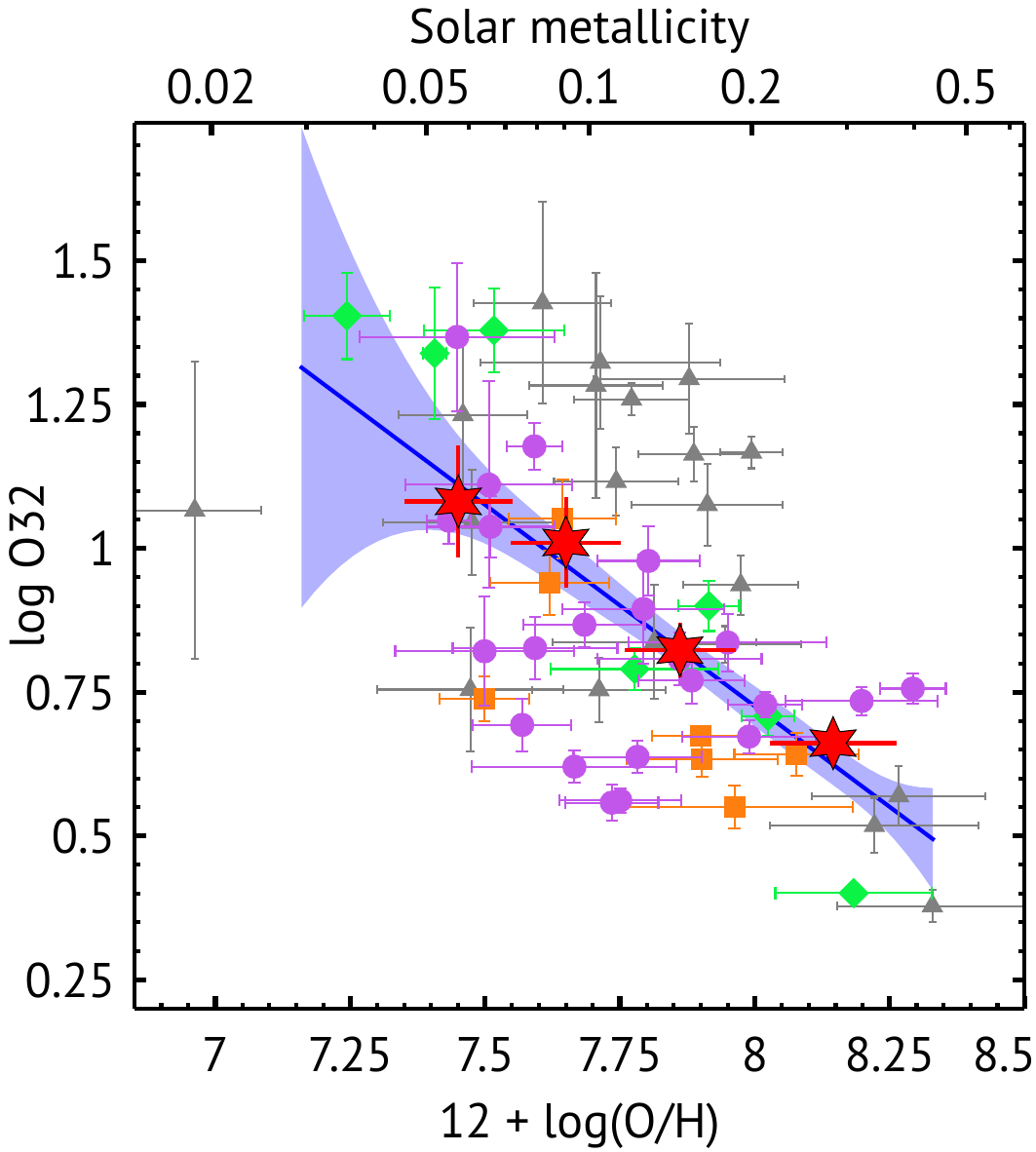}&
\includegraphics[width=0.33\textwidth]{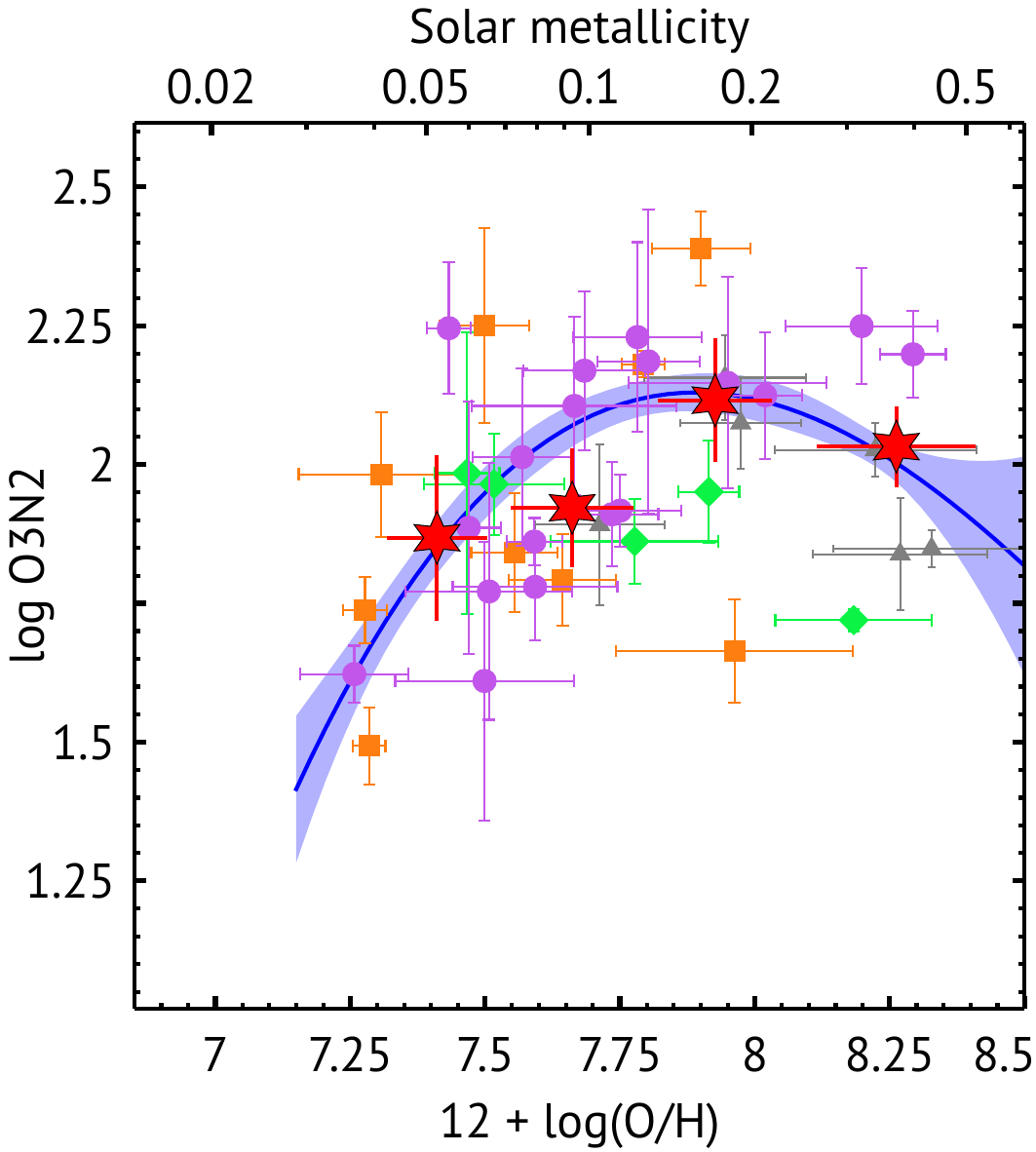}\\
\end{tabular}
    \caption{The relationship between
    $T_e$-based metallicity and 
    strong-line ratios 
    (R3, R2, R23, O32, Ne3O2, and O3N2)
    is shown for a sample of 67
    individual galaxies within the
    redshift range $3 < z < 10$.
    Purple and green data points
    represent galaxies from the 
    GOODS-N and GOODS-S fields, 
    respectively, 
    based on the JADES data release 3 
    \citep{2024arXiv240406531D}.
    Orange data points correspond 
    to galaxies from the 
    PRIMAL survey 
    \citep{2024arXiv240402211H},
    while gray data points indicate 
    JWST-detected galaxies compiled 
    from the literature 
    \citep{2024ApJ...962...24S}.
    Large red data points denote
    the metallicity-averaged line ratios. 
    Blue curves represent our
    best-fit polynomials 
    with $1\sigma$ uncertainties 
    for each diagnostic.}
\label{fig:calibration_best_fit}
\end{figure*}

\begin{figure}
\centering
\includegraphics[width=0.5\textwidth]{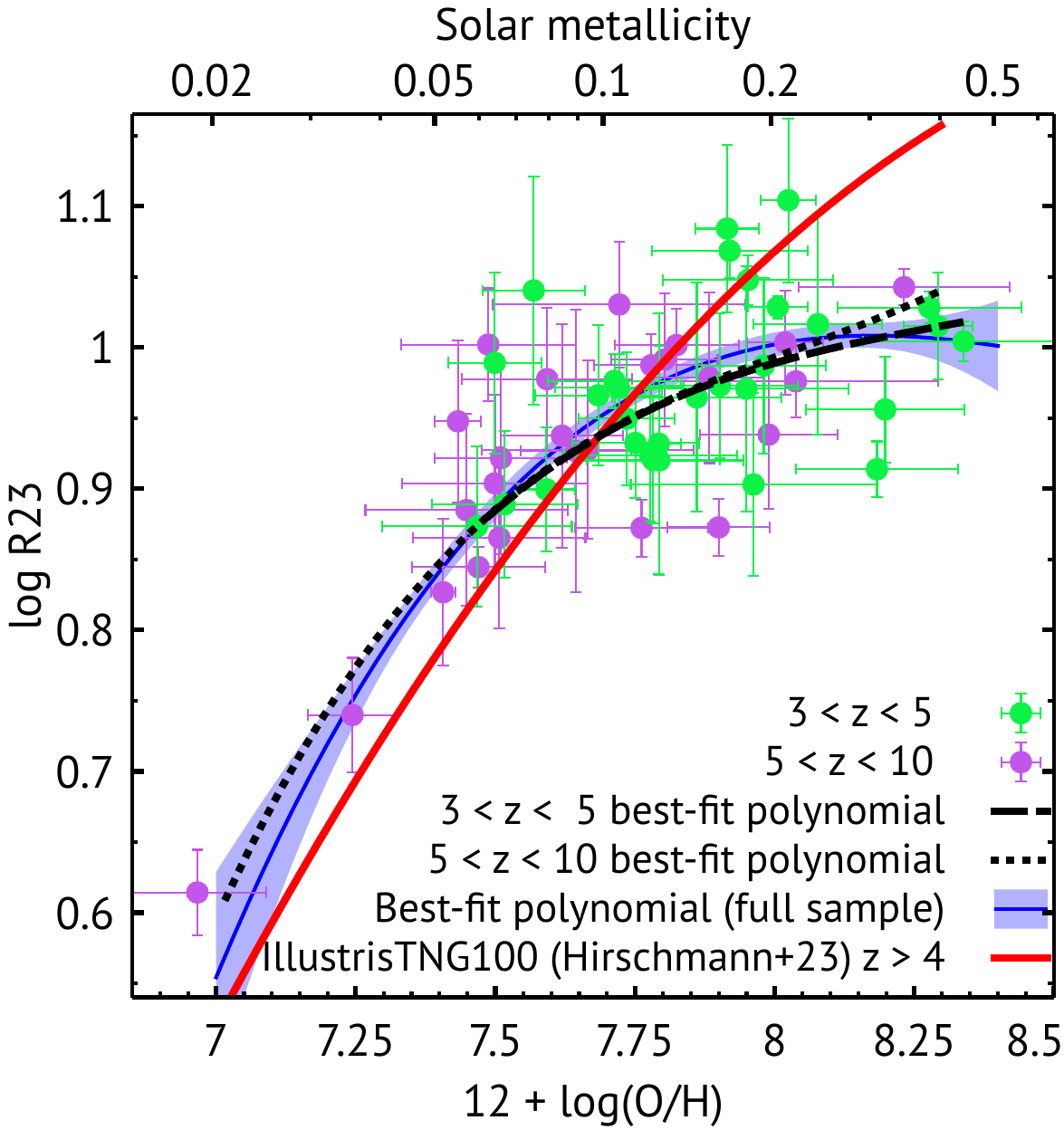}
    \caption{The $R23$ 
    line-ratio vs. metallicity
    is shown for two subsamples:
    galaxies within 
    $3 < z < 5$ (green circles) and 
    galaxies within $5 < z < 10$ 
    (purple circles).
    The black dashed curve 
    represents the best-fit
    metallicity calibration for
    the $3 < z < 5$ subsample,
    while the black dotted 
    curve shows the best-fit
    calibration for the 
    $5 < z < 10$ subsample. 
    The blue curve represents 
    the overall metallicity 
    calibration for the full sample, 
    covering $3 < z < 10$, with
    1$\sigma$ uncertainty.
    The red solid curve displays the 
    $R23$ vs. metallicity 
    relationship derived from 
    $z > 4$ IllustriTNG 100 galaxies, 
    adapted from 
    \citet{2023MNRAS.526.3504H}.
   }
\label{fig:rcap_redshift}
\end{figure}

\section{Metallicity diagnostic for strong lines at high-redshift}

\subsection{Gas-phase Metallicity}\label{sec:calibrations}
{ We utilize a sample of 
67 galaxies within 
the redshift range of $3 < z < 10$ (42
new galaxies and 25 from
literature) to
determine $T_e$-based 
metallicities. }
This compiled dataset 
significantly expands the 
current sample of 
high-redshift galaxies,
incorporating 2.6 times 
the number used in previous 
JWST studies for calibrating 
$T_e$-based metallicities and
empirical diagnostics 
(e.g., \cite{2024ApJ...962...24S}) 
within $3 < z < 10$.
In Figure 
\ref{fig:calibration_best_fit},
we display the relationships
between key strong-line
metallicity indicators 
(outlined in Section 
\ref{sec:intro}) 
and the gas-phase oxygen 
abundance for the
entire dataset based on the
direct $T_e$ method.
To estimate the average 
strong-line ratios, we
binned the dataset into four bins 
7.2--7.5, 7.5--7.8,
7.8--8.1, and $>$ 8.1, 
as shown with
big red stars in Figure 
\ref{fig:calibration_best_fit}.

To establish our new calibrations,
we adopt a polynomial fitting approach,  
characterized by the 
following  functional form:

\begin{equation}\label{eq:calibration_rel}
    {\rm log}(R) = \sum_{n=0}^{N} c_n \ x^n,
\end{equation}
where $R$ represents the line ratio under consideration, and $x$ is defined as $x = 12 + \log(\text{O/H}) - 8.0$ 
\citep{2001ApJ...556L..63A}. 
The $c_n$'s are the fit coefficients. 
Figure 
\ref{fig:calibration_best_fit} 
shows the best-fit polynomials 
with $1\sigma$ uncertainties 
alongside the observed line 
ratios. 
The best-fit coefficients
are listed in
Table \ref{tab:calibrators}.

\begin{table*}
   \caption{Best-fit values of the metallicity calibrators shown in Figure \ref{fig:calibration_best_fit}.}
    \centering
    \setlength{\tabcolsep}{5pt}
    \begin{tabular}{cccccccc}       
         Diagnostic & N$_{\rm gal}$ &c$_{0}$ & c$_{1}$ & c$_{2}$ & c$_{3}$ & RMS & $\sigma$\\ 
        \hline
        \hline
        R3 & 65 & 0.819 & $-$0.022 & $-$0.334 & 0.143 & 0.07 & 0.01 \\ 
        R2 & 59 & 0.049 & 0.41 & $-$0.20 & 0.80& 0.17 & 0.02  \\
        R23 & 59 & 1.0 & 0.074 & $-$0.245 & 0.128 & 0.067 & 0.004  \\
        O32 & 55 & 0.726 & $-$0.70 & -- & -- & 0.211 & 0.023 \\
        Ne3O2 & 54  & $-$0.46 & $-$0.75& 0.4 & -- & 0.194 & 0.020 \\
        O3N2 & 38 &2.12 & $-$0.22 & $-$0.94& 0.33 & 0.21 & 0.04 \\
        $\hat{\rm R}$ & 59 & 0.796 & 0.082 & $-$0.216 & 0.472 & 0.06 & 0.018  \\
        \hline     
    \end{tabular}
    \label{tab:calibrators}
\end{table*}

The R23 and R3 parameters are 
widely utilized diagnostics 
for deriving metallicity
\citep[e.g.,][]{1979MNRAS.189...95P,1984MNRAS.211..507E,2001A&A...369..594P,2001MNRAS.323..887C,2023ApJS..269...33N}. 
{ R23 ratio is long-known to 
have two metallicity solutions,
when metallicity approaches
12+log(O/H) $\geq$ 8.0
\citep[e.g,][]{1979MNRAS.189...95P,2002ApJS..142...35K}.}
Studies focusing on the local 
Universe (z $\sim$ 0) and 
local analogs 
\citep[e.g.,][]{2017MNRAS.465.1384C, 2018ApJ...859..175B, 2020MNRAS.491..944C, 2022ApJS..262....3N}, 
as well as prior high-redshift
($z > 3$) investigations, 
albeit with limited sample sizes 
\citep{2024A&A...681A..70L, 2024ApJ...962...24S}, 
show that R3
parameters can also yield two 
distinct metallicity solutions:
one corresponding to a low-metallicity 
regime and the other to a high-metallicity 
regime, 
with a turnover around 
$12 + \log(\mathrm{O/H}) \sim 8.0$. 
Our analysis of galaxies within the 
redshift range $3 < z < 10$
also exhibits a similar pattern, 
consistent with earlier studies, 
where R23 and R3  shows a 
decline at lower metallicities, 
particularly at 
$12 + \log(\mathrm{O/H}) < 7.7$  
\citep{2022ApJS..262....3N}.

{ R2 is typically not utilized 
as a standalone metric for
measuring metallicity,
since it is primarily
sensitive to the
ionization parameter in 
addition to metallicity}
\citep[e.g.,][]{2002MNRAS.330..876C,2004AJ....127.2002K,2019ARA&A..57..511K}.
R2 is used as a supplementary
tool to resolve degeneracies
associated with the R23 and R3 
calibrations. 
For R2, the turnover manifests
at 
$12 + \log(\mathrm{O/H}) \sim 8.5$
in the low-redshift Universe 
\citep{2020MNRAS.491..944C, 2022ApJS..262....3N}. 
Given that our sample is
constrained to the metallicity
range of
$12 + \log(\mathrm{O/H}) < 8.4$,
we detect an increasing trend 
in R2 with metallicity.

O32 is an indirect metallicity 
indicator that can show 
considerable scatter, 
affecting its precision,
as seen in Figure 
\ref{fig:calibration_best_fit}.
However, it is useful for 
distinguishing between 
metallicity solutions from
diagnostics like R23 
\citep[e.g.,][]{2002ApJS..142...35K, 2006A&A...459...85N, 2019A&ARv..27....3M,2023ApJS..269...33N}.  
We find a negative trend
between the O32 ratio and metallicity, 
showing no signs of reaching
a plateau, with a Spearman 
coefficient of $\rho_s = -0.47$
and a p-value of 0.001, 
indicating a weak correlation.

Ne3O2 index can be a good
metallicity indicator in 
the $\hii$ regions due to
its monotonic trend with 
metallicity and minimal 
sensitivity to reddening 
\citep{2007A&A...475..409S}. 
For Ne3O2, the Spearman 
correlation yields a coefficient of  
$\rho_s = -0.63$ with a p-value of 
$7.4\times10^{-7}$, indicating a 
statistically significant negative 
correlation, 
which can be used
to break the degeneracy between
two metallicity solutions from
R23 and R3 diagnostics.
We present, for the first time, 
diagnostic tools for the O3N2
line-ratio tailored to 
high-redshift galaxies.
Figure \ref{fig:calibration_best_fit} 
displays the O3N2 line-ratio
for our sample of galaxies. 
Our best-fit curve shows an
initial increase in line ratio
with metallicity, up to
\(12 + \log(\text{O/H}) \sim 7.9\), 
followed by a decline, 
resulting in doubly
degenerate metallicity values.

\subsection{Redshift evolution of strong line-ratio metallicity indicators}
In this section, we explore 
the
potential redshift
evolution of strong line-ratio 
metallicity indicators
 within 
$3 < z < 10$.
We only use the \(R{23}\) 
line-ratio for this purpose, 
as they demonstrate the
least scatter in the data
(see Figure \ref{fig:calibration_best_fit})
To facilitate this analysis, 
we divided our
entire sample of 
galaxies into two subsamples: 
one consisting of 36 galaxies
with redshifts between 
\(z = 3\) and \(z = 5\), 
and the other comprising 23
galaxies with redshifts 
between \(z = 5\) and \(z = 10\).
Figure \ref{fig:rcap_redshift} 
illustrates the redshift evolution
for the ranges \(z = 3 - 5\) and
\(z = 5 - 10\).
The black dashed line represents
the best-fit polynomial for 
the subsample within \(3 < z < 5\),
while the black dotted line 
represents the best-fit
polynomial for the subsample with 
\(5 < z < 10\). 
Figure
\ref{fig:rcap_redshift}
also
shows the modeled 
\(z > 4\) relation for 
IllustriTNG100
galaxies  from 
\citet{2023MNRAS.526.3504H}, 
shown as thick red line.

For \(R{23}\), 
the subsample with \(3 < z < 5\) 
shows metallicity differences of 
less than 0.02 dex compared to
the calibration in the 
\(5 < z < 10\) range,
indicating a negligible variation. 
This aligns with both simulations
and previous studies. 
\citet{2023MNRAS.526.3504H} 
used the TNG100 
galaxy population to study 
metallicity calibrations up 
to redshift 8 and observed minimal 
variation between 
\(z = 4\) and
\(z = 8\) in \(R{23}\). 
Similarly, 
\citet{2024ApJ...972..113G} 
found no significant variation 
in \(R{23}\) 
calibration between
\(z = 3\) and \(z = 5\) 
using mock emission line 
data from the SIMBA simulations. 
\citet{2024ApJ...962...24S} 
examined the residuals around 
the best-fit \(R{23}\) 
calibrations as a function
of redshift and did not
find significant trends,
implying limited redshift 
evolution up to \(z \sim 8\). 
While they highlighted the 
potential benefits of additional
\(T_e\) measurements for
refining the calibration
across this range,
our expanded sample
to approximately
three times larger within 
\(3 < z < 10\)--allows us
to divide the data into
two redshift bins. 
This larger sample size 
provides further evidence 
supporting the lack of 
significant redshift 
evolution in \(R{23}\).

\section{Discussion}

\subsection{Mass--Metallicity relation: ``direct''-$T_{e}$ based}
\begin{figure*}
    \centering
\includegraphics[width=0.6\textwidth]{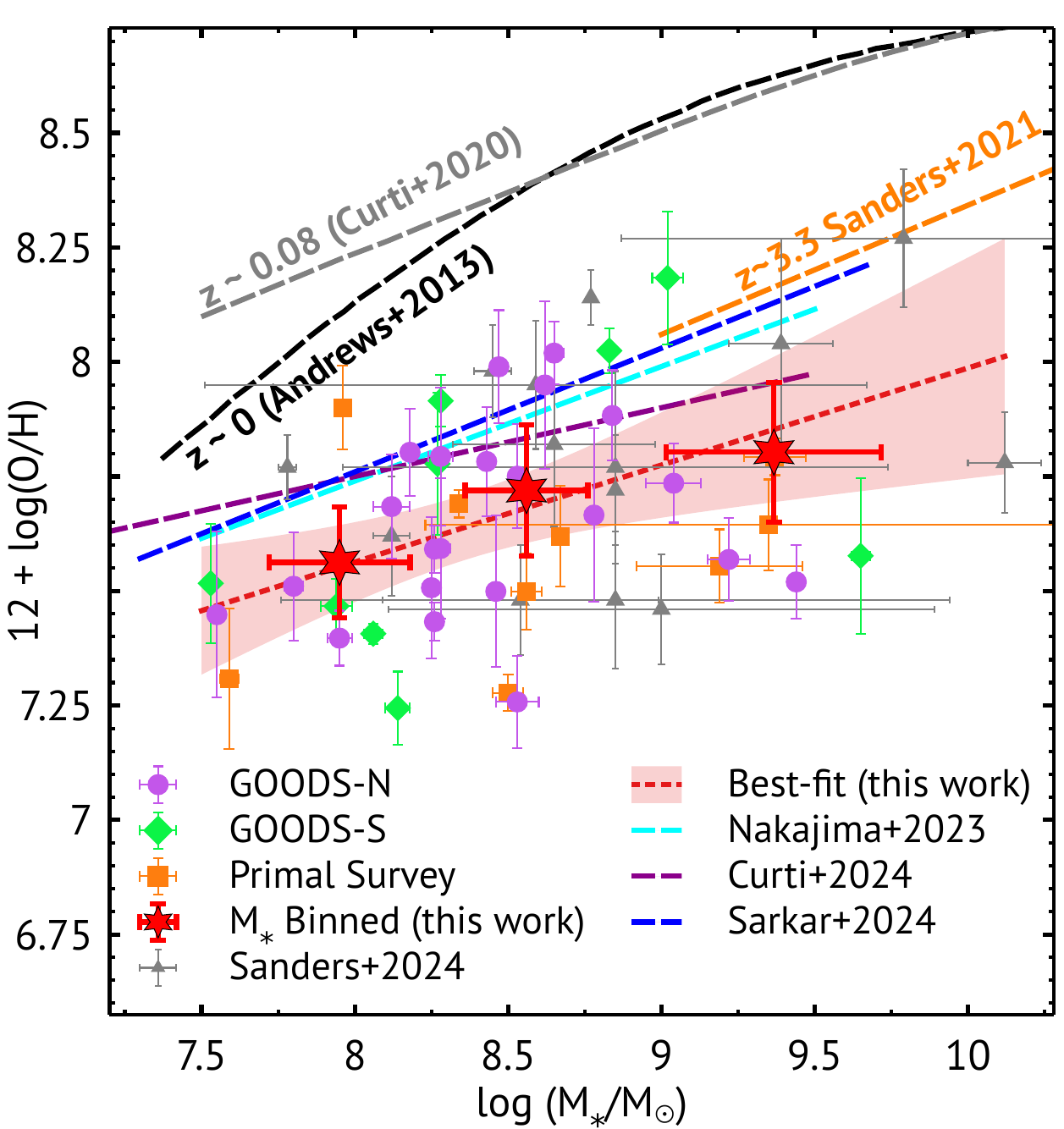}
    \caption{The mass-metallicity
    relation (MZR) is shown,
    where metallicity is derived
    using the direct-$T_e$
    based method. 
    Green diamonds, purple circles, 
    and orange squares represent 
    data from the GOODS-S, GOODS-N,
    and Primal survey, respectively.
    Gray triangles denote 
    a sample adapted from literature.
    Red stars show the stellar 
    mass-averaged metallicities.
    The red dashed curve illustrates
    our best-fit MZR with a 
    $1\sigma$ uncertainty, 
    with best-fit parameters 
    $\gamma = 0.211 \pm 0.120$ 
    and $Z_{10} = 7.986 \pm 0.205$. 
    For comparison, MZR curves 
    from other calibration-based high-redshift 
    JWST studies by 
    \citet{2024A&A...684A..75C}, 
    \citet{2023ApJS..269...33N}, and 
    \citet{2024arXiv240807974S} are 
    included, as well as MZR curves from 
    low-redshift studies by 
    \citet{2020MNRAS.491..944C}, 
    \citet{2013ApJ...765..140A}, and 
    \citet{2021ApJ...914...19S}.
    }
    \label{fig:MZR}
\end{figure*}

Characterizing the scaling relation
between stellar mass and gas-phase 
metallicity in galaxies, 
known as the mass–metallicity relation (MZR) 
\citep[e.g.,][]{1979A&A....80..155L,2004ApJ...613..898T,2006ApJ...647..970L,2024arXiv240706254G,2024arXiv240807974S}, 
is essential for understanding the
processes that regulate the 
growth of early galaxies. 
This relation reflects the 
interplay between gas accretion,
star formation, metal enrichment, 
and outflows that shape 
the baryon cycle. 
Here, we present, for the first time, 
the MZR for galaxies in the 
redshift range $3 < z < 10$, 
with metallicities derived from 
the direct-$T_e$ method and 
stellar masses obtained from 
continuum fitting, 
as detailed in Sections 
\ref{sec:prism_spectra_fitting} 
and \ref{sec:metal_measure}. 
{ We re-scaled our stellar mass 
measurements
to a common \citet{2003PASP..115..763C} 
IMF.}
Figure \ref{fig:MZR} shows 
the MZR for our complete sample of
67 galaxies, including 25 from 
literature.

To compare the slope
of the our MZR relation 
with the previous studies,
we fit the measured average
MZR with the relation, as 
described in several previous 
studies \citep[e.g.,][]{2021ApJ...914...19S,2024arXiv240702575C,2023ApJS..269...33N,2024arXiv240807974S},
\begin{equation}\label{eq:mzr}
    12 + \log(\text{O/H}) = \gamma \times \log\left(\frac{M_\star}{10^{10} M_\odot}\right) + Z_{10},
\end{equation}
where the slope $\gamma$ and offset $Z_{10}$
(the gas-phase metallicity 
at a stellar mass of $10^{10} M_\odot$)
are derived by fitting to the observed
MZR. To estimate the average MZR,
we divided our sample into 
three stellar mass bins: 
log($M_{\ast}/M_{\odot}$) = 7.5–-8.25, 8.25–-9, and $>$ 9, 
ensuring that each bin contains
at least 15 galaxies.
We determined a best-fit slope 
of $\gamma = 0.211 \pm 0.120$ and 
a metallicity intercept of
$Z_{10} = 7.986 \pm 0.205$. 
The average points and the 
best-fit regression line, 
along with their 
1$\sigma$ uncertainties, 
are shown in Figure \ref{fig:MZR}.

{This study presents the first mass-metallicity relation (MZR) at high redshift  derived using the \(T_e\)-based method with \textit{JWST}. Prior studies such as JADES + Primal \citep{2024arXiv240807974S}, CEERS \citep{2023ApJS..269...33N}, and JADES \citep{2024A&A...684A..75C} have employed calibration-based techniques to estimate metallicities.}
Table \ref{tab:mzr_comp} provides 
a comparison of our best-fit
slope ($\gamma$) and metallicity
intercept ($Z_{10}$) with those from previous studies.
While the slope and normalization for
our sample are consistent within 1$\sigma$ uncertainty with other high-redshift studies that used local calibrations, we find that at a stellar mass of $M_{\ast} = 10^8 , M_{\odot}$, our best-fit curve lies approximately 0.2 dex lower than those derived using calibration-based methods.  The methodological distinction between $T_e$-based and calibration-based approaches may account for the systematic offset observed in metallicity measurements between our work and these studies.

We also compared our results with previous MZR studies conducted at lower redshifts using the $T_e$-based method. The metallicity at a given stellar
mass in our sample is noticeably 
lower than the metallicity 
curves at $z \sim 0$ from 
\citet{2013ApJ...765..140A} 
and $z \sim 0.08$ from 
\citet{2020MNRAS.491..944C}. 
This difference varies with 
stellar mass, showing an 
offset of about 0.6 dex at
$10^8 \, M_{\odot}$ and 
0.7 dex at $10^9 \, M_{\odot}$, 
consistent with previous 
high-redshift galaxy studies 
\citep[e.g.,][]{2024A&A...684A..75C,2023ApJS..269...33N,2024arXiv240807974S}.
Additionally, we find that our 
direct $T_e$-based metallicity 
is approximately 0.3 dex lower 
than the $z \sim 3.3$ MZR 
curve reported by 
\citet{2021ApJ...914...19S} 
for galaxies with stellar
masses of $10^9 \, M_{\odot}$.

Recent JWST studies reveal 
considerable intrinsic scatter 
in the properties of
high-redshift galaxies.
However, performing a statistically
robust assessment of this scatter
within scaling relations remains 
challenging due to the 
limited size of our sample.
We determine the intrinsic scatter following the methodology outlined in \citet{2024arXiv240807974S}:
\begin{equation}\label{eq:scatter}
    \sigma_{\rm scatter} = \sqrt{\sigma_{\rm obs}^2 - \sigma_{\rm measured}^2},
\end{equation}
where \(\sigma_{\text{obs}}\) 
represents the observed scatter
in the full sample
around the best-fit MZR line, 
and \(\sigma_{\text{measured}}\) 
indicates the average
uncertainty in the 
metallicity measurement.
We estimate \(\sigma_{\text{scatter}}\) 
for our entire galaxy sample to
be approximately 0.09 dex. 
This value aligns with the 
intrinsic scatter of 0.08 dex 
reported for low-stellar-mass 
galaxies (log \(M_\star / M_\odot \lesssim 9\))
at redshifts \(3 < z < 10\) 
\citep{2024A&A...684A..75C}. 
However, it is lower than the
intrinsic scatter of 0.16 dex 
observed in JADES + Primal 
survey galaxies 
\citep{2024arXiv240807974S} and
the 0.16–0.18 dex range found in 
dwarf galaxies at redshifts 
\(z = 2\) to \(3\) with stellar
masses between \(10^8\) 
and \(10^9 \, M_\odot\) 
\citep{2023ApJ...955L..18L}.

\begin{table}
   \caption{Comparing MZR with 
   different studies following 
   equation \ref{eq:mzr}. 
   $Z_{8}$ is converted to 
   $Z_{10}$ using $Z_{10} = 2\gamma + Z_{8}$.}
    \centering
    \setlength{\tabcolsep}{5pt}
    \begin{tabular}{llcc}
        
        Study & $z$ range & $\gamma$ & $Z_{10}$ \\
        \hline
        This work (direct) & 3--10 & 0.21 $\pm$ 0.03 & 7.99 $\pm$ 0.21 \\
        \hline
        \citet{2024arXiv240807974S} & 4--10 & 0.27 $\pm$ 0.02 & 8.28 $\pm$ 0.08 \\
        \hline
        \citet{2023ApJS..269...33N} & 4--10 & 0.25 $\pm$ 0.03 & 8.24 $\pm$ 0.05 \\
       \hline
       \citet{curti2023jadesinsightslowmassend} & 3--10 & 0.17 $\pm$ 0.03 & 8.06 $\pm$ 0.18 \\
        \hline
        \citet{2021ApJ...914...19S} & 0 & 0.28 $\pm$ 0.01 & 8.77 $\pm$ 0.01 \\
       & 2.3 & 0.30 $\pm$ 0.02 & 8.51 $\pm$ 0.02 \\
       & 3.3 & 0.29 $\pm$ 0.02 & 8.41 $\pm$ 0.03 \\
        \hline
        \citet{2023ApJ...955L..18L} &  2 & 0.16 $\pm$ 0.02 & 8.50 $\pm$ 0.13 \\
       & 3 & 0.16 $\pm$ 0.01 & 8.40 $\pm$ 0.06 \\
        \hline
        \citet{2023NatAs...7.1517H} & 7--10 & 0.33 & 7.95 \\
        \hline
        \citet{2024ApJ...960L..13H} & 1.9 & 0.23 $\pm$ 0.03 & 8.54 $\pm$ 0.12 \\
       & 2.88 & 0.26 $\pm$ 0.04 & 8.57 $\pm$ 0.15 \\
        \hline
    \end{tabular}
    \label{tab:mzr_comp}
\end{table}

\subsection{Metallicity calibrations at z$\sim$0}

\begin{figure*}
\centering
\begin{tabular}{ccc}
\includegraphics[width=0.33\textwidth]{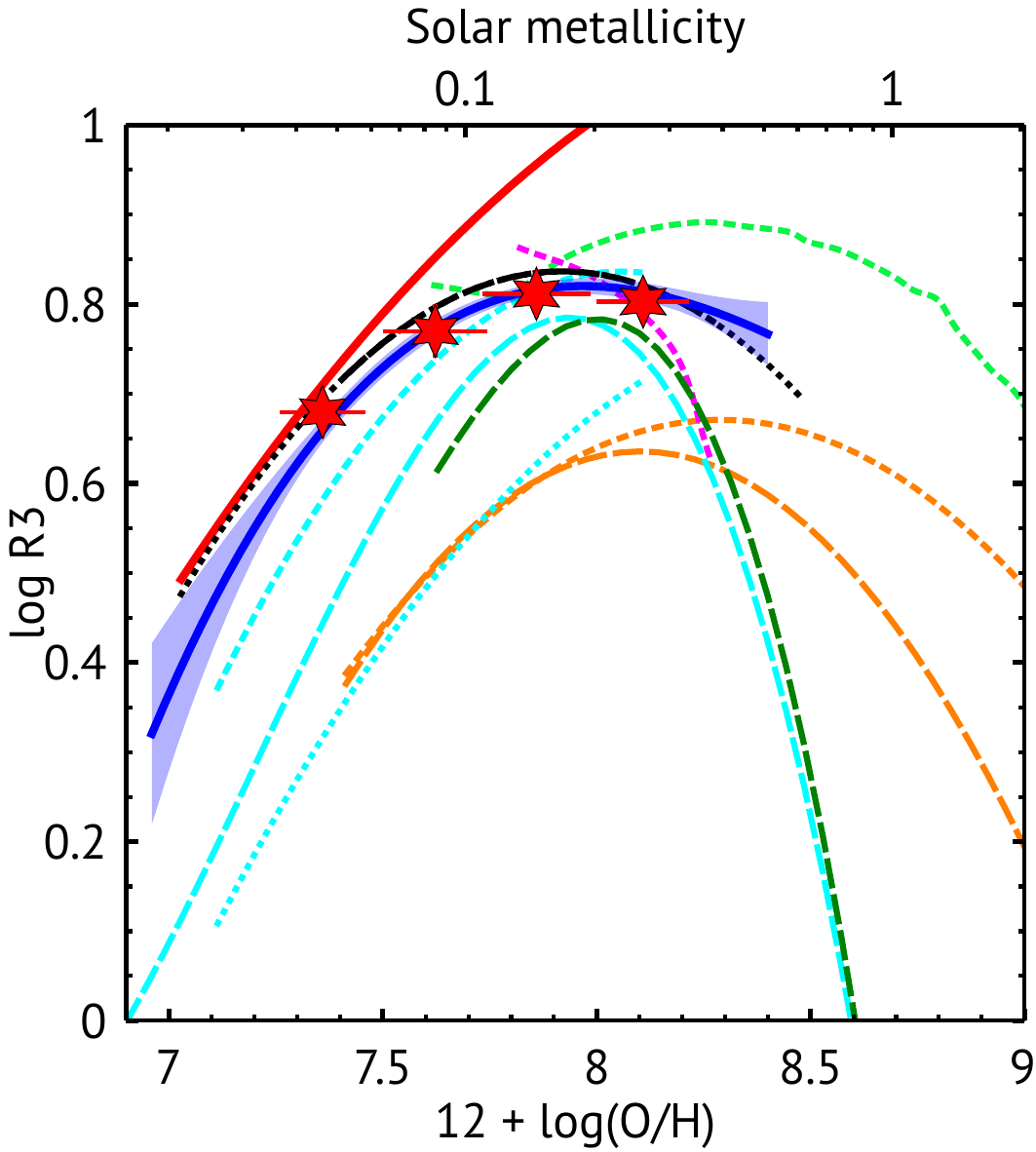} &   \includegraphics[width=0.33\textwidth]{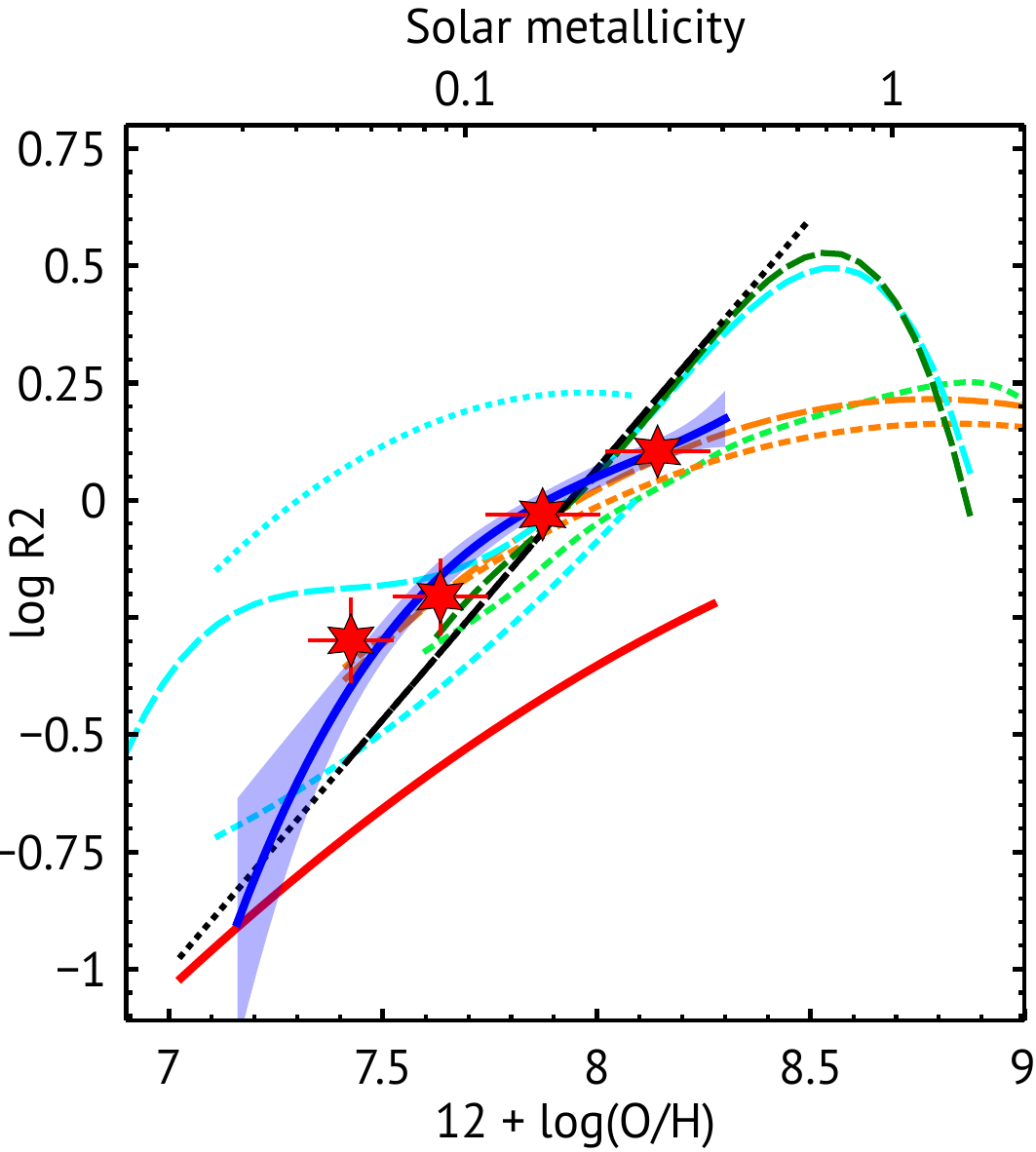}&\includegraphics[width=0.33\textwidth]{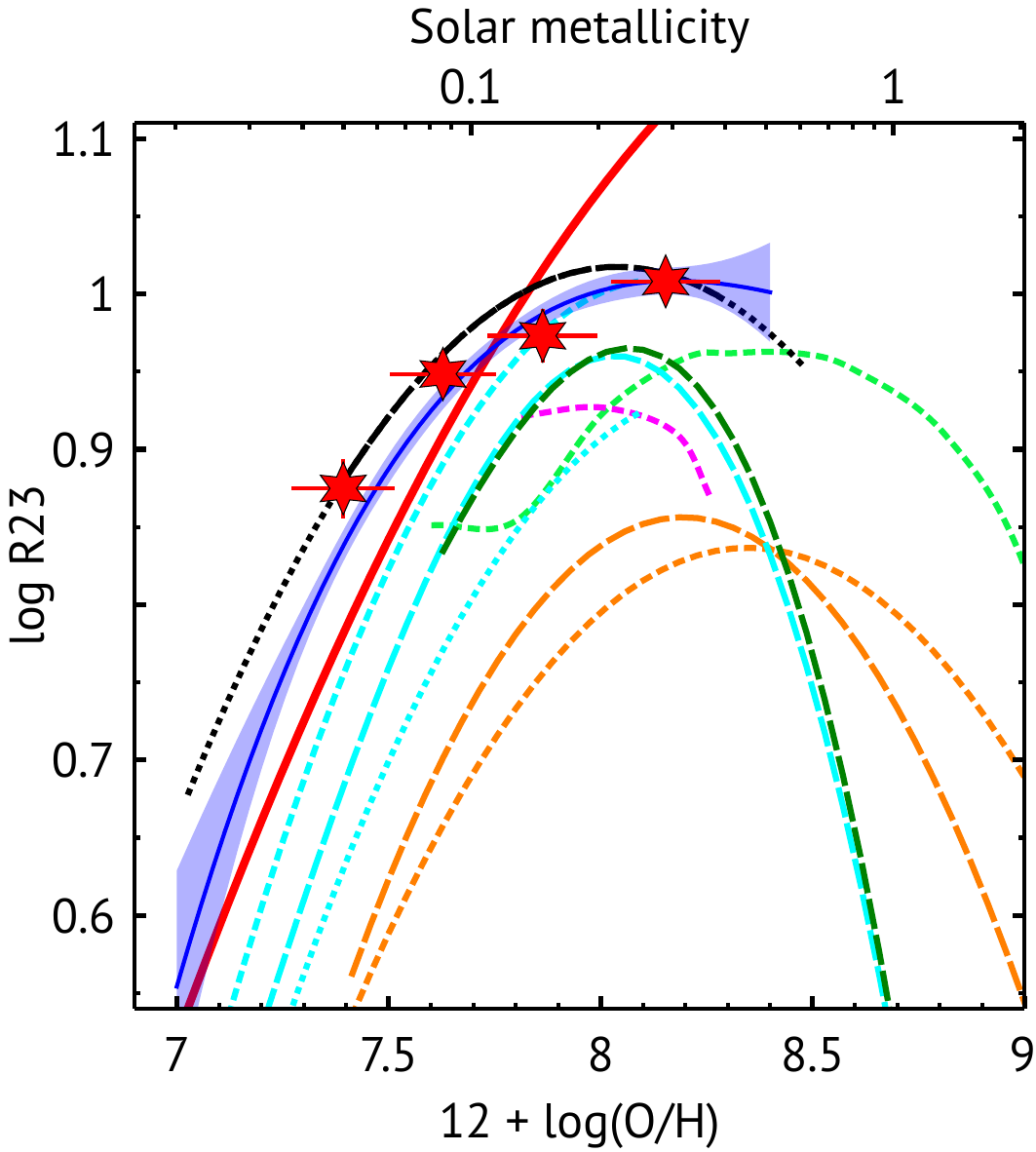}\\
\end{tabular}  
\begin{tabular}{ccc}
\includegraphics[width=0.33\textwidth]{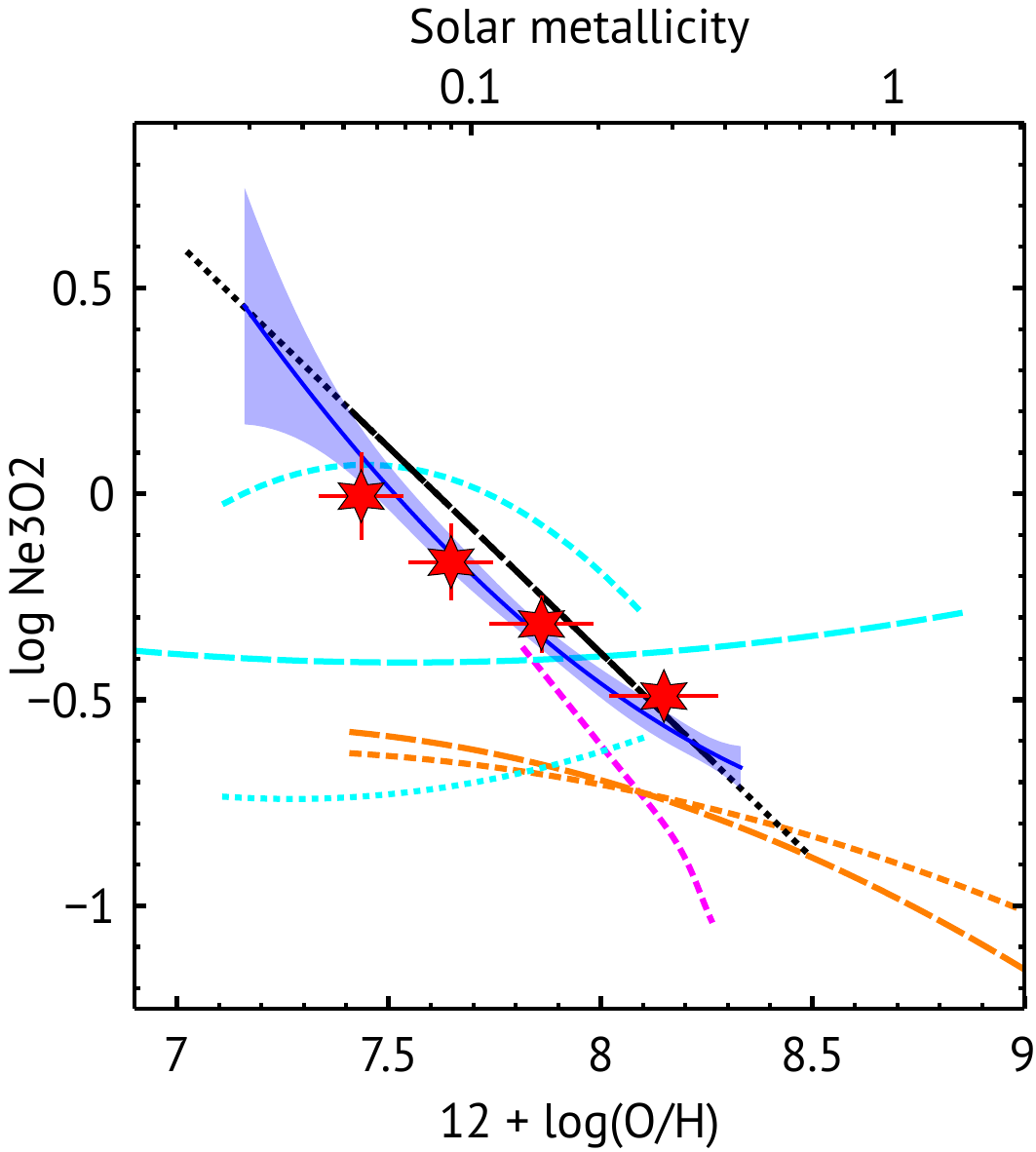} &   \includegraphics[width=0.33\textwidth]{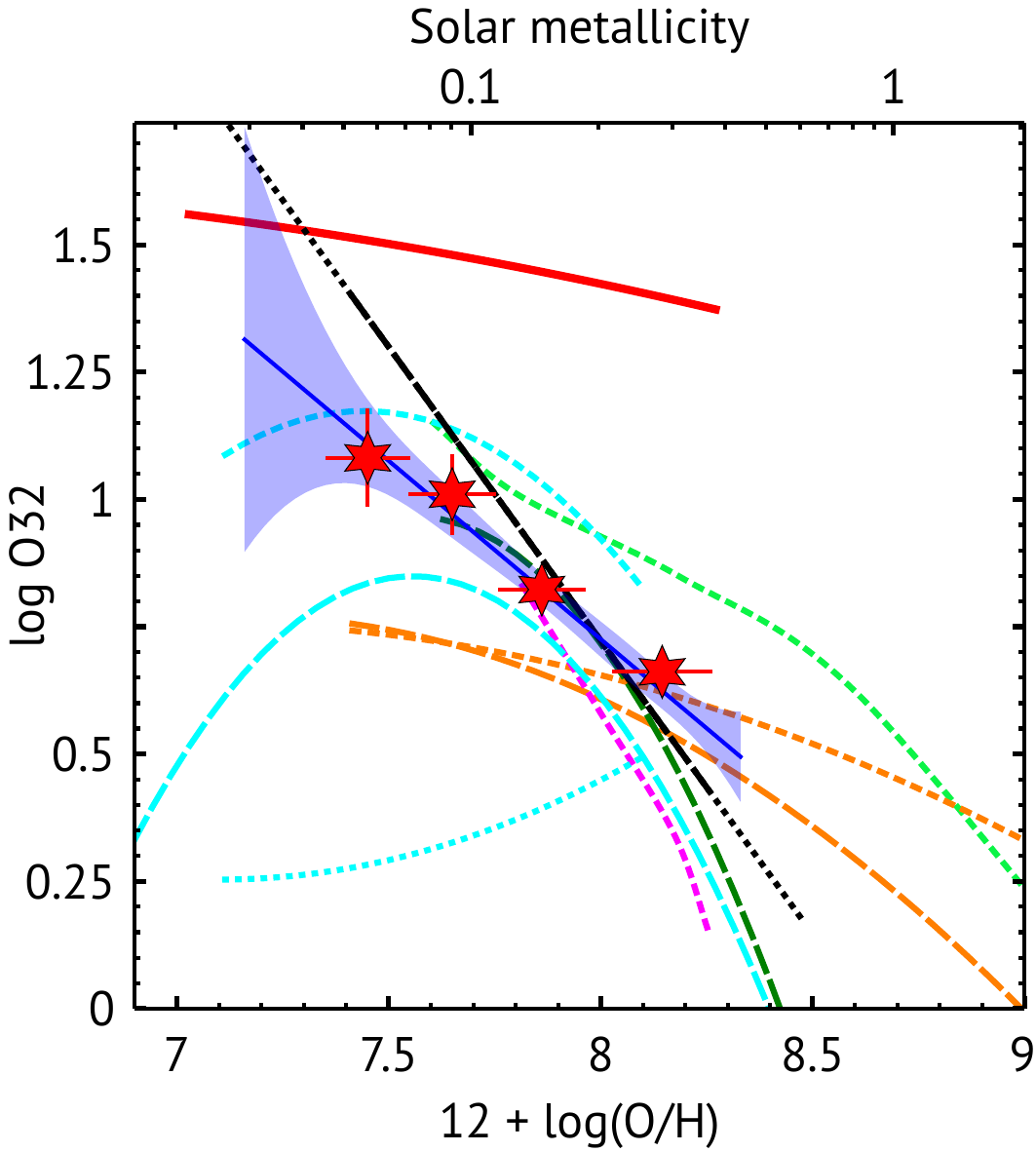}&\includegraphics[width=0.33\textwidth]{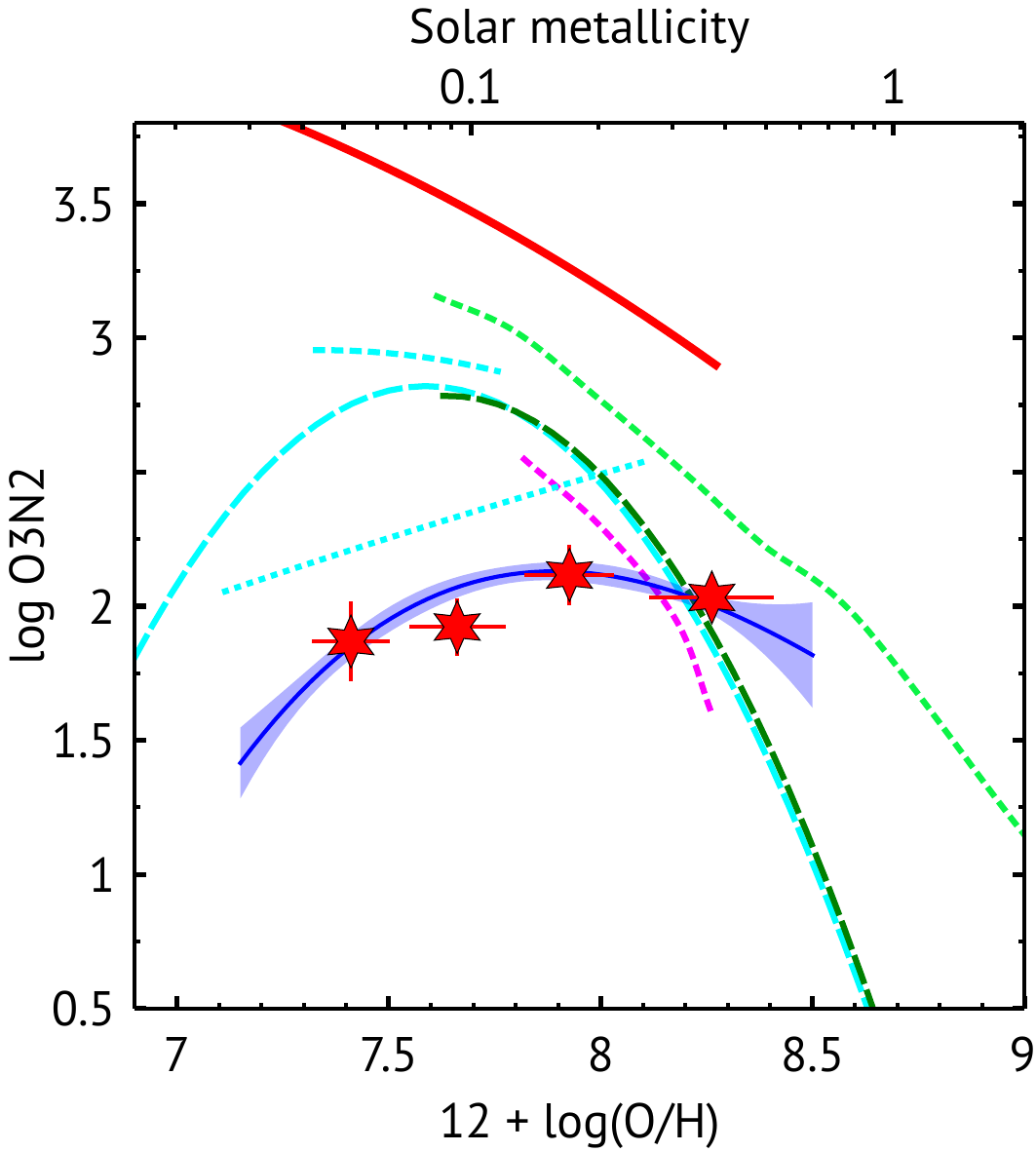}\\
\end{tabular}
\begin{tabular}{ccc}
 & \hspace{50pt}\includegraphics[width=0.7\textwidth]{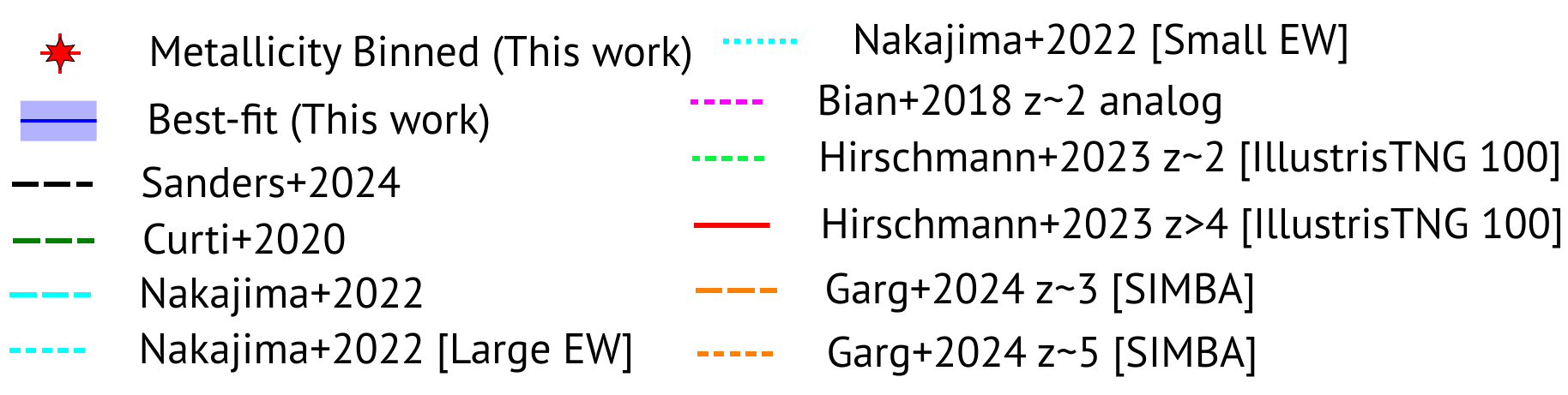} &\\
\end{tabular}
    \caption{ Comparison of our strong-line metallicity calibration for high-redshift galaxies in the range $3 < z < 10$ (blue curve with shaded regions showing the 1$\sigma$ uncertainty) with previous observational calibrations and simulations, as color-coded in the legend. The comparison includes strong-line calibrations for $z \sim 0$ from \citet{2015ApJ...813..126J}, \citet{2020MNRAS.491..944C}, \citet{2022ApJS..262....3N} along with local analogs that occupy similar regions to $z \sim 2$ star-forming galaxies on the Baldwin-Phillips-Terlevich (BPT) diagram \citep{2018ApJ...859..175B}.  High-redshift calibrations for galaxies in the range $2 < z < 9$ are also shown \citep{2024ApJ...962...24S}, alongside strong-line metallicity calibrations derived from simulations, including the SIMBA cosmological galaxy formation model at $z \sim 3$ and $z \sim 5$  \citep{2024ApJ...972..113G}, and the TNG100 galaxy population simulation at $z \sim 2$ and $z > 4$ \citep{2023MNRAS.526.3504H}.}
\label{fig:calibration_best_fit_comp}
\end{figure*}

Figure 
\ref{fig:calibration_best_fit_comp} 
presents a comparison between
our high-redshift strong-line
calibrators and those derived 
from local metallicity
calibrators at z$\sim$ 0 
\citep[e.g.,][]{2020MNRAS.491..944C, 2021ApJ...914...19S, 2022ApJS..262....3N}. 
The R3 and R23 diagnostics
exhibit consistently larger
line ratios at a given 
metallicity compared to
local calibrations.
Both R3 and R23 show a 
turnover around
$12 + \log(\text{O/H}) \sim 8.0$, 
which is consistent with the
turnover observed in the 
local R3 and R23 diagnostics. 
For R3, the high-redshift
calibrations either 
underpredict or overpredict
metallicity by $\sim$ 
0.2--0.4 dex compared to the
z $\sim$ 0 calibrators, 
depending on whether they 
lie on the left or right
side of the turnover. 
In a similar manner, 
the high-redshift calibrations
for R23 exhibit metallicity 
differences 
(either positive or negative) 
of around 0.1--0.3 dex when 
compared to the z$\sim$0 
calibrators.

The R3 ratio at high redshift
largely aligns with the 
z$\sim$0 calibrations  
within the metallicity range
$7.7 < 12 + \log(\text{O/H}) < 8.4$, 
exhibiting marginally higher
line ratios at a given 
metallicity compared to z$\sim$ 0.
Conversely, for $12 + \log(\text{O/H}) < 7.7$, 
the R2 ratio at high 
redshift is consistently 
lower than that of the 
local calibrations at 
the same metallicity. 
At a fixed O/H, high-redshift
calibrations show elevated
Ne3O2 ratio compared to 
local  Ne3O2 diagnostics. 

The O32 ratio, 
an ionization-sensitive 
diagnostic commonly employed
as a degeneracy breaker in
the low-z universe, 
consistently exhibits higher
values at the same metallicity
in high-z galaxies compared
to local O32 calibrations. 
However, the substantial 
scatter observed in the 
high-z O32 ratios—spanning 
up to an order of 
magnitude--results in a 
weak correlation between
O32 and metallicity 
(see Section 
\ref{sec:calibrations}), 
raising concerns about 
the reliability of O32 
as a diagnostic tool for 
resolving degeneracies in 
the high-z Universe. 
This scatter has been 
previously attributed to the 
significant diversity in 
ISM conditions observed
within a sample of 25 
high-redshift galaxies,
and our sample, 
which is approximately
$> 2.5$ times larger, 
further reinforces this 
interpretation. 

{

The calibrations derived based on Te metallicities in $\hii$ regions in the local universe has been found to be subject to systematic discrepancies \citep{2019ARA&A..57..511K}. The underlying factors contributing to the observed discrepancies are complex and challenging to unravel. These discrepancies stem from sample  biases, the assumptions made during the estimation of metallicity, and several other factors discussed in Section \ref{sec:caveats}, resulting in variations of up to 1 dex in metallicity estimates (see Figure 8 in \citet{2019ARA&A..57..511K}). The sample biases introduced by these differences may be attributed to assumptions that 
$\hii$ regions will always display the auroral lines if observed for an adequate duration and that $\hii$ regions showing auroral lines are representative of all $\hii$ regions and galaxies.
}

Our capacity to evaluate
metallicity diagnostics for
values above 
$12 + \log(\text{O/H}) > 8.4$ 
is presently constrained by 
the absence of data points 
beyond this range.
The apparent absence of a 
turnover in R2 may be
attributed to the lack of
data points at 
$12 + \log(\text{O/H}) > 8.4$. 
To better understand the 
behavior of high-redshift 
metallicity calibrations 
at higher metallicities 
($12 + \log(\text{O/H}) > 8.4$), 
a more comprehensive analysis, 
incorporating a larger high-z 
[$\oiii$]$\lambda$4363 sample 
and stacked spectra from 
hundreds of galaxies, is needed.

\subsection{High-redshift metallicity calibrations}

Before the JWST era, several efforts
aimed to expand the sample of 
high-$z$ galaxies with robust oxygen
abundance measurements using the
direct method.
This was often achieved by stacking
spectra of local analogs of 
high-$z$ star-forming galaxies to
develop calibrations consistent 
with the high-$z$ regime 
\citep{2015ApJ...813..126J, 2018ApJ...859..175B, 2021MNRAS.504.1237P} 
or by obtaining direct metallicity
measurements from the
[$\oiii$]$\lambda4363$ auroral line 
\citep{2020MNRAS.491.1427S}.

High-redshift metallicity calibrations
were initially explored using a 
sample of 46 star-forming galaxies
covering a redshift range of 
$z = 1.4$ to $z = 8.7$,
including 25 JWST-detected 
galaxies at $z > 2$ \citep{2024ApJ...962...24S}. 
This study provided $T_{e}$-based 
calibrations for strong emission 
lines such as 
$R{23}$, $R{2}$, $R{3}$, $O{32}$, 
and $Ne3O{2}$. 
A subsequent study by 
\citet{2024A&A...681A..70L} 
generally found agreement with 
these calibrations for most 
strong emission lines, 
though they identified some 
deviations for $R{2}$ and $O32$, 
which were attributed to significant
scatter in the data. 
To enhance the robustness of 
these calibrations, 
\citet{2024A&A...681A..70L} 
emphasized the importance of 
obtaining larger samples of 
[$\oiii$]$\lambda4363$ detections
for the high-redshift universe.

We revisited the $T_e$-based 
metallicities using a sample of 67
[$\oiii$]$\lambda4363$ lines at $z > 3$,
incorporating 42 new 
detections from our analysis. 
The solid blue lines in 
Figure \ref{fig:calibration_best_fit_comp}
represent our calibration,
which is compared to the
calibrations of 
\citet{2024ApJ...962...24S}, 
shown with black lines.  
For $R{23}$ and $R{3}$, 
the expanded sample indicates 
a good agreement with the 
calibrations of 
\citet{2024ApJ...962...24S}, 
with the metallicity offset 
being $\lesssim 0.05$ dex.  
The $Ne3O2$ calibrator also 
shows reasonable agreement, 
with metallicity offsets of 
$\lesssim 0.07$ dex.
For $R{2}$, 
while there is generally 
reasonable alignment, 
there is a maximum metallicity
offset of around $\sim 0.15$ dex.
Whereas \citet{2024ApJ...962...24S} 
suggest a linear relationship for
the metallicity calibrator, 
our findings indicate a potential
non-linear trend,
as shown in Figure 
\ref{fig:calibration_best_fit} 
and 
\ref{fig:calibration_best_fit_comp}.
For \( O_{32} \), we note a slight difference in the slopes of the metallicity calibrators between our calculations and those presented by \citet{2024ApJ...962...24S}, with the maximum metallicity offset reaching approximately \(\sim 0.4\) dex.

\subsection{New metallicity calibrators}
The large intrinsic scatter in
individual line-flux 
ratios and the $T_e$--based metallicity
for high-redshift galaxies makes 
the calibration a 
challenging task. 
\citet{2024A&A...681A..70L} proposed
a novel metallicity calibration relation,
$\hat{R}$ based on observations
from 465 low-metallicity objects compiled
from several previous observations.
$\hat{R}$ is defined as 
\begin{equation}\label{eq:rhat}
    \hat{R} = {\rm cos}(\phi)\ {\rm log}(R2) + {\rm sin}(\phi)\ {\rm log}(R3),
\end{equation}
which is equivalent to a rotation 
of the R2-R3 plane around the
O/H axis. 
The process to find $\phi$ 
involves fitting a fourth-order polynomial
to the resulting
$\hat{R}$ ratio versus the metallicity, 
in the form of $\hat{R} = \sum_{n=0}^{N} c_n x^n$ where
$x$ = 12 + log(O/H) $-$ 8.0, 
and the angle $\phi$ 
minimizes the scatter in
metallicity from the best-fit
relation.
\citet{2024A&A...681A..70L} found a best-fit
$\phi$ of 61.82 deg, which gives
$\hat{R}$ = 0.47 log($R2$) + 0.88 log($R3$). 

\begin{figure*}
    \centering
\begin{tabular}{ccc}
\includegraphics[width=0.5\textwidth]{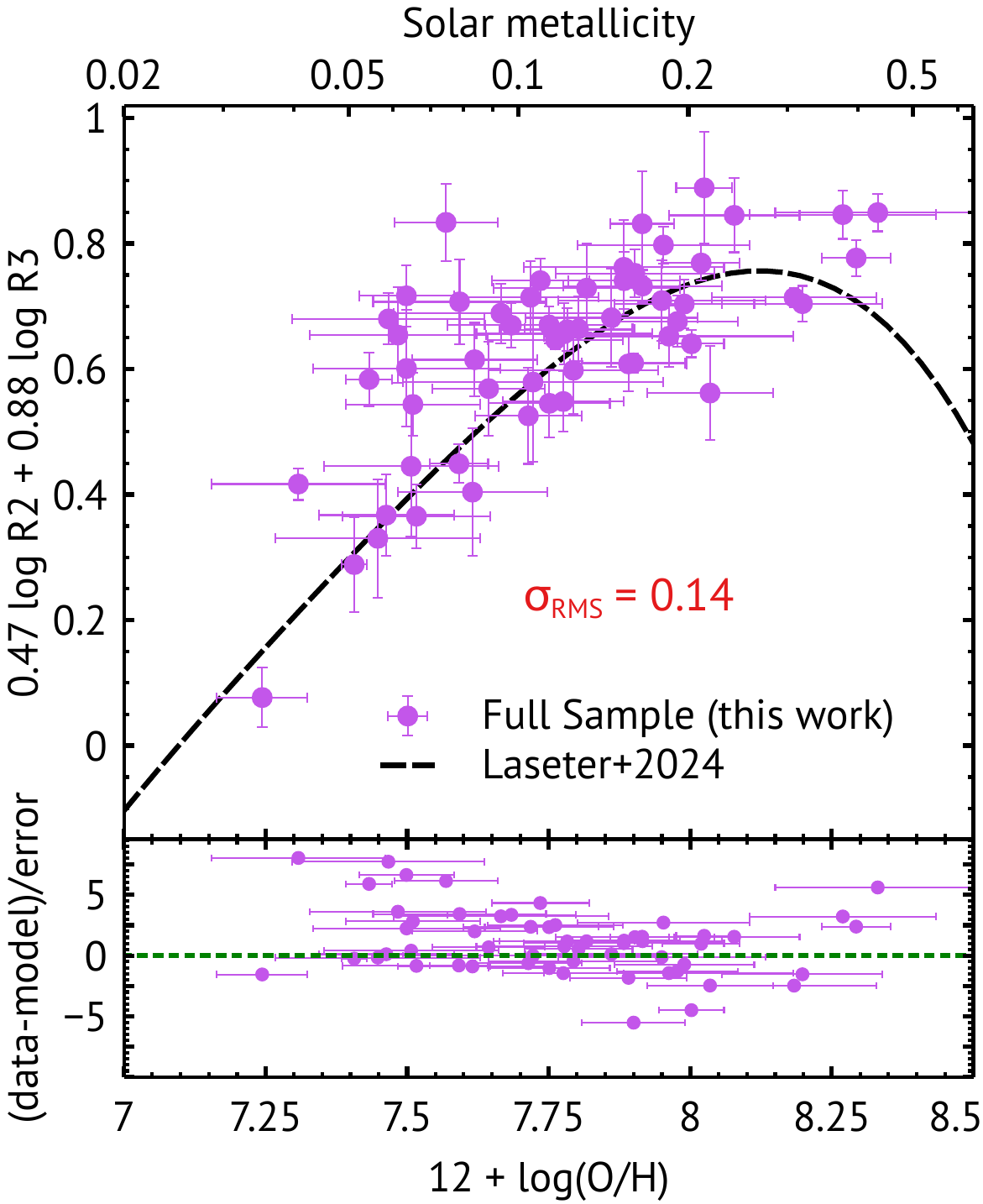} &   \includegraphics[width=0.5\textwidth]{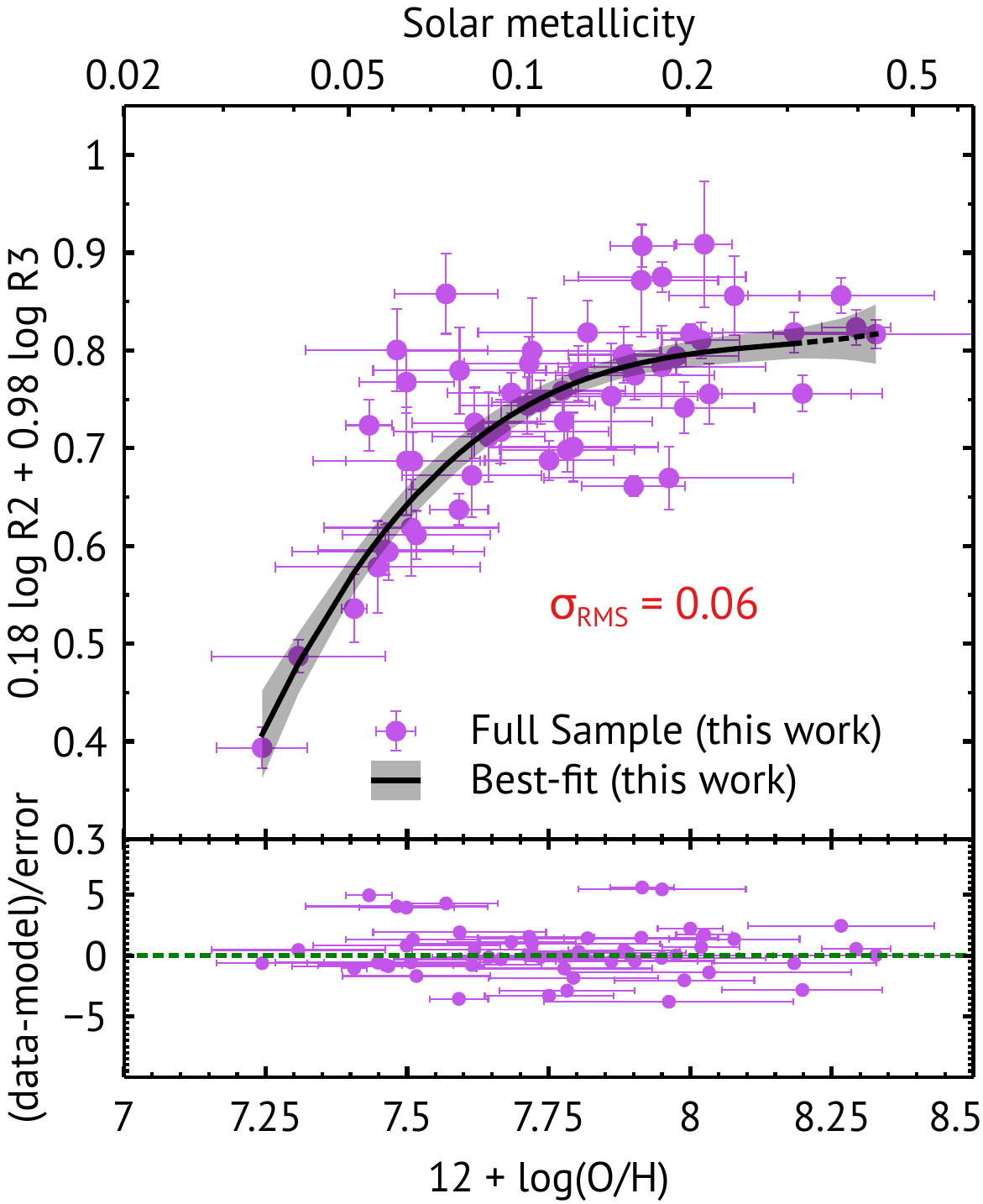}\\
\end{tabular}  
    \caption{The novel metallicity 
    calibration relation, $\hat{R}$, 
    is displayed. 
    In both panels, 
    purple circles represent our 
    entire sample. 
    {\it Left:}
    $\hat{R}$ derived using 
    the best-fit relation, 
    $\hat{R} = 0.47 \, \text{log} \, R2 
    + 0.88 \, \text{log} \, R3$,
    from \citet{2024A&A...681A..70L}. 
    Their best-fit polynomial 
    curve is indicated by the 
    dashed line.
    {\it Right:}  
    $\hat{R}$ derived exclusively 
    from our high-redshift sample,
    using the relation 
    $\hat{R} = 0.18 \, \text{log} \, R2 
    + 0.98 \, \text{log} \, R3$. 
    The best-fit polynomial curve
    for this calibration is shown 
    as a black solid line and dashed
    line with $1\sigma$ 
    uncertainties. 
    { The solid line represents the best-fit within the range where the statistical significance is higher, supported by a larger number of data points.}
    Bottom panels in both figures
    illustrates the $\Delta\chi$ values. { Our new calibration substantially reduces the intrinsic scatter in the data, offering a more reliable tool for measuring metallicity at higher redshifts.}
    }
    \label{fig:rcap_fits}
\end{figure*}

In this paper, we tested this new 
metallicity calibration for our full
sample of galaxies within $3 < z < 10$.
Figure \ref{fig:rcap_fits} illustrates
the estimated $\hat{R}$ of our full
sample alongwith best-fit curve predicted
by \citet{2024A&A...681A..70L}.
We next follow the same prescription
as \citet{2024A&A...681A..70L}
to constrain the $\phi$ value for our
high-redshift sample. 
We found  
$\phi$ = 79.8 $\pm$ 2.4 deg, which gives
\begin{equation}
    \hat{R} = 0.18\ {\rm log}(R2) +
    0.98\ {\rm log}(R3).
\end{equation}
Figure \ref{fig:rcap_fits} illustrates 
this new calibration relation for our
full sample with the best-fit curve.
The best-fit coefficients for  
our new $\hat{R}$ calibration
\{0.796, 0.082, $-$0.216, 0.472\} are listed
in Table \ref{tab:mzr_comp}.
To quatify the scattering in both
$\hat{R}$ relations, we measure
RMS scatter ($\sigma_{\rm RMS}$) of derived
$\hat{R}$ from their respective
best-fit curve.
For \citet{2024A&A...681A..70L},
we find $\sigma_{\rm RMS}$ = 0.14 dex
and for our new $\hat{R}$ we find 
$\sigma_{\rm RMS}$ = 0.06 dex.

Our new $\hat{R}$ calibration 
significantly reduces the scatter
in the observed line-ratio vs. metallicity,
as clearly seen in Figure \ref{fig:rcap_fits}, and,
therefore can be used to estimate
metallicity in high-redshift galaxies.
However,
we caution that our new $\hat{R}$
calibration
is derived from a limited sample of
59 high-redshift galaxies.
{ Users should apply these
new calibration
within the range 
of 7.2 $\leq$ 12 + log(O/H)
$\leq$ 8.2 to
ensure reliable
metallicity estimates.} 
Though our new $\hat{R}$
calibration hints towards the challenges
of using low-redshift metallicity 
calibrators in 
deriving metallicity in
high-redshift galaxies, 
more high quality data 
is needed to put stronger constraints
on the high-redshift metallicity 
calibrators.

\begin{figure*}
    \centering
\begin{tabular}{cc}
\includegraphics[width=0.5\textwidth]{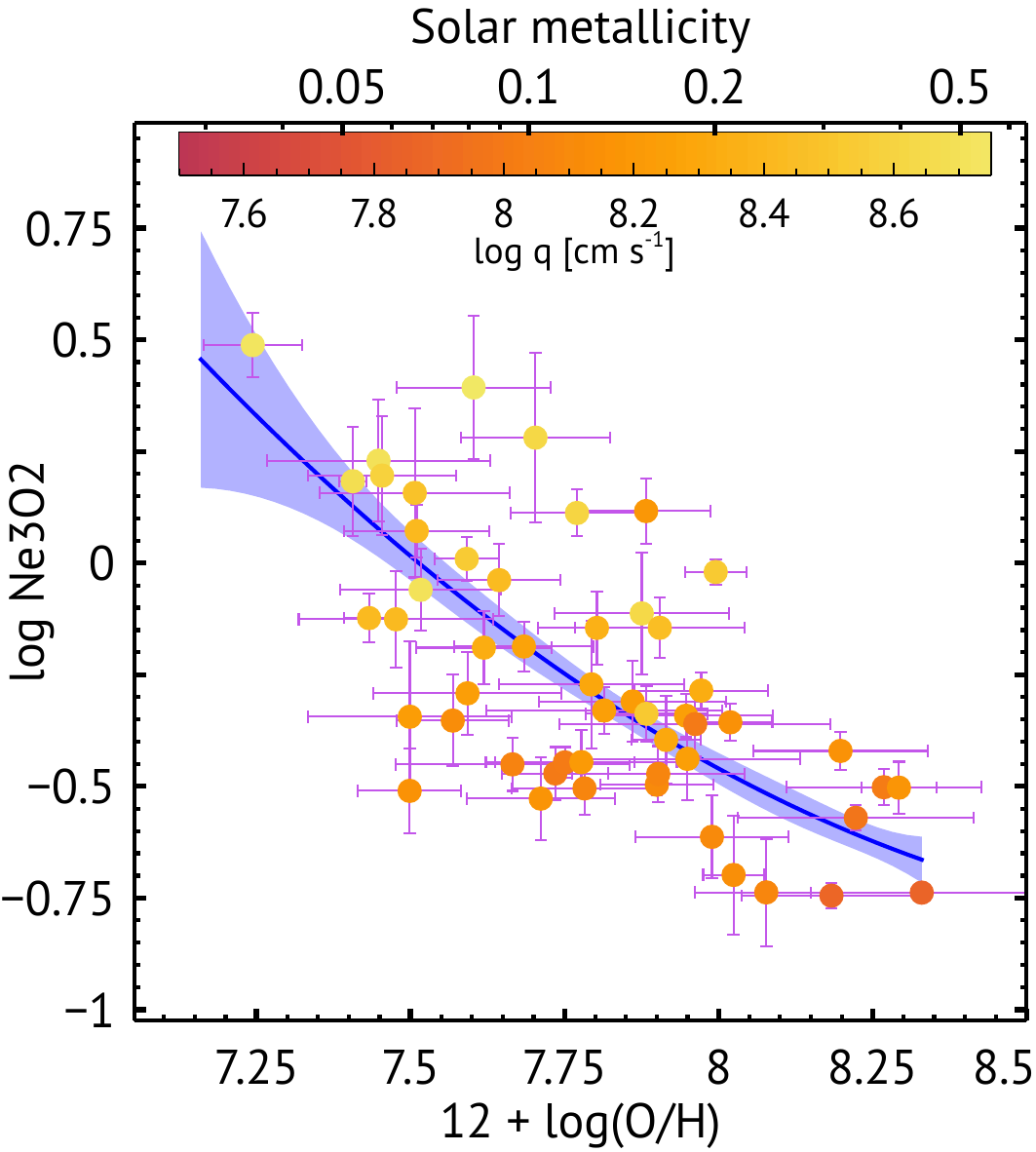} &   \includegraphics[width=0.5\textwidth]{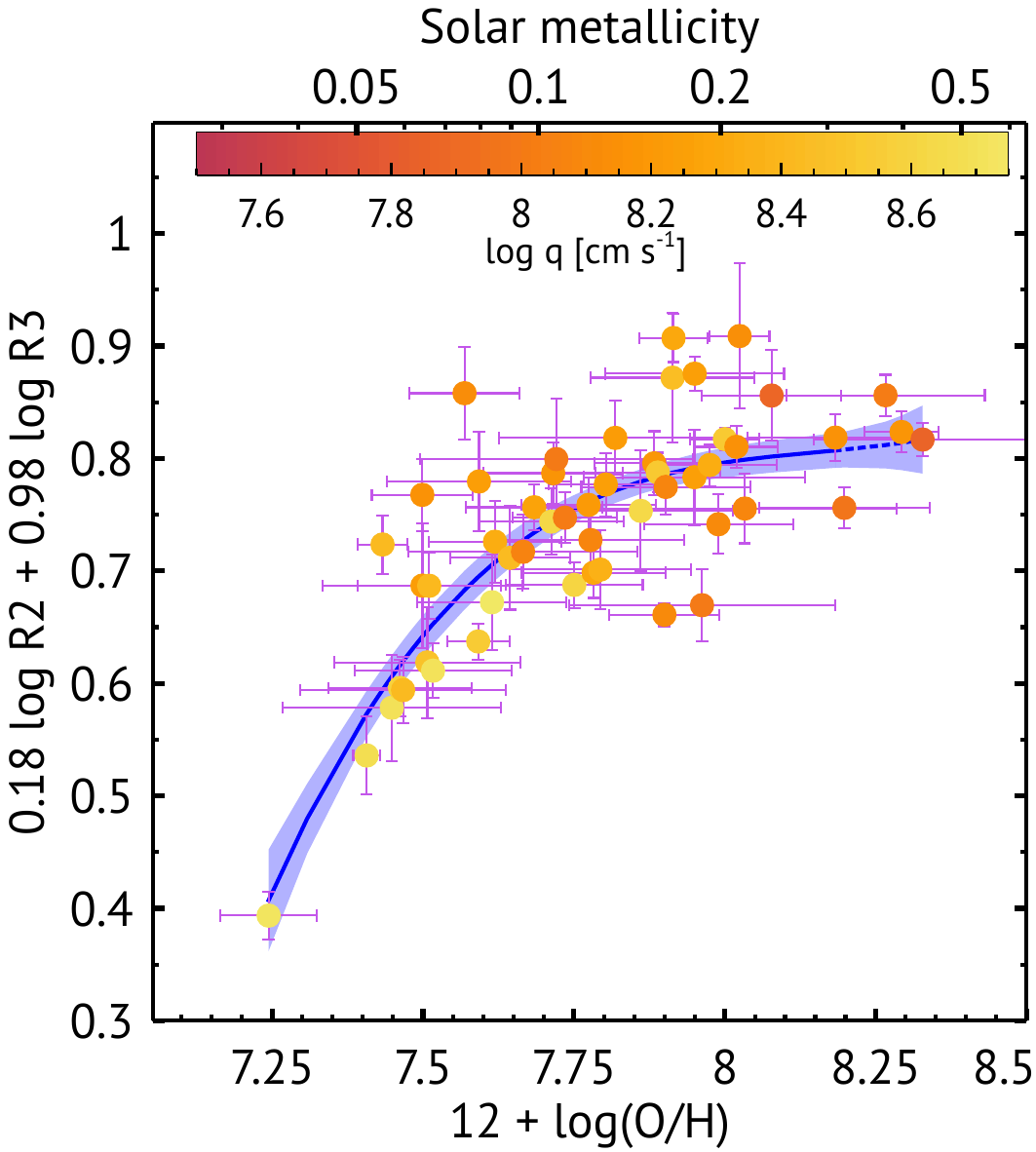}\\
\end{tabular}  
    \caption{{\it Left:} 
    The Ne3O2 line ratio versus metallicity is shown, with derived ionization parameters for individual galaxies based on the O32 line ratio. Ionization parameters are represented as a color gradient, with the colorbar displayed at the top. The best-fit polynomial for the entire sample (as listed in Table \ref{tab:calibrators}) is shown as a blue solid line, along with the
    1$\sigma$ uncertainty.
    The ionization parameter varies by $\sim$ 1 dex for a given metallicity around 7.6, indicating that Ne3O2 diagnostics may be affected by the
    log $q$--log(O/H) relation.
    {\it Right:} Similar
    to the {\it Left} panel
    but for the $\hat{R}$ calibration.
    The $\hat{R}$ diagnostic is less
    impacted by the log $q$--log(O/H) relation, demonstrating the robustness of this calibration relation.}
    \label{fig:ionization}
\end{figure*}

{ 
\subsection{Caveats in $T_e$-based metallicity and calibrations}\label{sec:caveats}

The $T_e$-based method 
offers a relatively direct 
approach to measuring
gas-phase metallicity 
compared to estimates 
derived from local
calibration relations.
However, it is not without
limitations and can suffer 
from significant systematic 
uncertainties due to various
factors
\citep{2019ARA&A..57..511K}.
One major challenge lies 
in the simplified multi-zone
models used to describe the 
temperature structure within 
ionized nebulae.
These models often lack
critical information about
the temperature of the 
O$^+$ zone, 
as the required auroral
emission line doublet at
$[\oii]\lambda\lambda$7322,7332 is
frequently too faint to
detect or falls outside 
the wavelength range of
NIRSpec. 
As a result, the O$^+$/H ratio,
which can represent a substantial 
fraction ($\sim$ 22--30\%) 
of the total 
oxygen abundance,
must be inferred indirectly.
This is typically done 
using a proxy temperature,
such as $T_e(\text{N II})$, 
or through empirical 
relations linking 
\(T_e(\text{O II})\) to 
\(T_e(\text{O III})\).
Additionally, the 
$[\oiii]\lambda$4363 auroral line, 
which is critical for 
\(T_e\)-based metallicity
estimates, 
is predominantly produced
in hot,
metal-poor gas 
\citep[e.g.,][]{1967ApJ...150..825P,2008ApJ...681.1183K}. 
In metal-rich regions, 
the gas temperature is lower, 
and collisional excitation of
the [$\oiii$] lines is
less efficient. 
This leads to significant 
temperature gradients and
fluctuations within $\hii$ regions, 
which systematically
bias \(T_e\) measurements
to higher values and
result in underestimates
of metallicity 
\citep[e.g.,][]{2005A&A...434..507S,2020A&A...634A.107Y}.

Accurate determinations of 
electron temperature (\(T_e\))
require the detection of
multiple faint
emission lines, such as 
[$\oii$], [$\sii$], and [$\nii$]. 
These lines must be measured
with an intensity accuracy
better than 5\% relative to
Balmer recombination lines
\citep{2006MNRAS.372..293H}. 
More deep observations of
high-redshift galaxies with
JWST/NIRSpec are needed
to address these 
challenges and improve 
the reliability of 
\(T_e\)-based metallicity
measurements.

Another significant source
of systematic 
uncertainty in metallicity
estimates stems from the 
assumptions made in the
calculation of the 
ionization correction factor,
particularly for unobserved
ionization stages
such as $\oiv$.
Standard ICF calculations 
typically assume a 
uniform electron
temperature and a
simplified $\hii$
region composition, 
often considering only 
a single or a small 
subset of ionic species.
However, electron temperatures
can vary significantly--by as
much as 2000-–3000 K--across 
different $\hii$ regions
\citep[e.g.,][]{2006MNRAS.372..293H,2019ARA&A..57..511K}. 
$\hii$ regions are complex, 
multi-zone environments 
containing a wide variety of
atomic and ionic species. 
The ionization states of 
elements are strongly
influenced by the stellar
radiation field, 
which itself depends on 
detailed stellar atmosphere models.
In the absence of accurate data,
the assumed ICF can deviate 
substantially from the true 
distribution of unobserved
ionization stages, particularly 
in low-excitation $\hii$ regions. 
This deviation can result in 
significant underestimation of metallicity, 
with errors reaching up
to 0.2 dex when using the 
\(T_e\)-based method
\citep{2008MNRAS.383..209H}. 
Such uncertainties highlight
the need for improved models
and future JWST
observations to refine 
ICF calculations and enhance
the reliability 
of metallicity estimates.

The metallicity calibration 
diagnostics for certain 
line ratios, such as R2, O32, 
and Ne3O2, are known to
sensitive to variations
in the ionization
parameter
\citep[e.g.,][]{2024A&A...681A..70L,2008ApJ...681.1183K}. 
To test this  
for our sample of galaxies, 
we derived the 
ionization parameter, 
defined as the ratio of 
hydrogen-ionizing photon 
flux to the hydrogen
density in 
the $\hii$ region,
using observed 
O32 line ratios. 
We adopted the linear
relation between log(O32) 
and the ionization
parameter from 
\citet{2022ApJ...937...22P}:
\begin{equation}
{\rm log}\ q = (0.86 \pm 0.07)\ {\rm log\ O32} + (7.53 \pm 0.02),
\end{equation}
where $q$ represents
the ionization parameter
in units of 
cm s$^{-1}$.
Figure \ref{fig:ionization} 
illustrates the 
derived ionization 
parameters as a color 
gradient in 
Ne3O2 vs. metallicity 
and $\hat{R}$
vs. metallicity plots.
For the Ne3O2 ratio,
the ionization parameter 
varies by $\sim$ 1 dex 
for a given metallicity
in the range 7.5--7.8, 
indicating that the Ne3O2 
diagnostic may be affected by 
the log $q$--log(O/H) 
relation.
However, the limited 
statistical power of our 
sample restricts a more 
detailed investigation
of this dependency,
such as splitting the
entire sample into two 
or more \( q \)-bins. 
A more robust dataset 
with higher statistical 
power is necessary for such an 
analysis. 
Future JWST surveys are 
expected to provide larger
high-redshift galaxy samples 
with high-quality spectra, 
enabling a more 
comprehensive study.
In contrast, $\hat{R}$ 
diagnostic shows weak 
dependence on the ionization 
parameter, highlighting its 
robustness as a metallicity 
calibration diagnostic
for high-redshift galaxies.
}

\section{Summary}
We report the discovery of
[$\oiii$] auroral lines ([$\oiii$]$\lambda$4363) 
in 42 galaxies observed with 
JWST/NIRSpec within the redshift 
range $3 < z < 10$. 
Combined with previous observations
of 25 galaxies featuring [$\oiii$] 
auroral lines within the
same redshift range, 
this combined dataset of
67 galaxies significantly expands 
the sample of high-redshift
galaxies available for study,
nearly tripling the number 
used in prior JWST studies for 
calibrating $T_e$-based 
metallicities and empirical
diagnostics for $3 < z < 10$.
Our main findings are summarized below.

\begin{itemize}
    \item The 42 galaxies in 
    this study were selected from
    a sample within the redshift 
    range $3 < z < 10$, 
    primarily derived from 
    publicly available data releases, 
    including JADES/GOODS-S, 
    JADES/GOODS-N, and JWST-PRIMAL. 
    These galaxies were observed 
    using the NIRSpec low-resolution 
    PRISM/CLEAR configuration as 
    well as medium-resolution gratings. 
    Stellar masses were determined
    through spectral fitting of the
    PRISM spectra using BAGPIPES, 
    with the sample spanning a mass
    range of 
    $10^{7.5}$–$10^{10} \, M_{\odot}$.

    \item To measure emission line fluxes, 
    we utilized the medium-resolution
    grating spectra for each galaxy,
    focusing on key nebular emission
    lines when detected, 
    including the hydrogen Balmer lines, 
    [$\oii$]$\lambda\lambda$3727,3729, 
    [$\neiii$]$\lambda$3867, 
    [$\oiii$]$\lambda$4363, 
    [$\oiii$]$\lambda\lambda$5007,4959, 
    and [$\nii$]$\lambda$6584. 
    The line fluxes were corrected 
    for dust extinction using the 
    curve from \citet{2000ApJ...533..682C}.
    To distinguish star-forming 
    galaxies from AGN-dominated ones, 
    we applied the
    mass-excitation (MEx) diagram.

\item The electron temperatures 
in the high-ionization regions,
$T_{\rm e}$([$\oiii$]), 
were derived from the flux ratio of
[$\oiii$]$\lambda$5007 to 
[$\oiii$]$\lambda$4363, 
assuming an electron density 
of 300 cm$^{-3}$. 
For our sample of 42 galaxies,
$T_{\rm e}$([$\oiii$]) ranges from
12,000 to 24,000 K, consistent
with extremely metal-poor local
galaxies and previous JWST surveys.
Among these, 10 galaxies also show 
[$\oii$]$\lambda\lambda$7322,32 auroral
lines with a signal-to-noise ratio 
of S/N $\geq$ 2, allowing us to use the 
[$\oii$]$\lambda\lambda$3727,3729 to 
[$\oii$]$\lambda\lambda$7322,32 
line ratios to determine the 
electron temperature in the
low-ionization region, 
$T_{\rm e}$([$\oii$]). 
Electron densities for the 
low-ionization region were 
calculated using the 
[$\sii$]$\lambda\lambda$6716,31 doublet ratios. 
For these 10 galaxies, 
$T_{\rm e}$([$\oii$]) ranges 
between 10,830 and 20,000 K. 
We established a best-fit relation 
between $T_{\rm e}$([$\oii$]) and 
$T_{\rm e}$([$\oiii$]): $T_{\rm e}$([$\oii$]) = 
(0.58 $\pm$ 0.19) $\times$ $T_{\rm e}$([$\oiii$]) + 
(4520 $\pm$ 2000) K, 
which aligns with observations 
of local galaxies. 
Oxygen abundances were calculated 
for our sample using {\tt PyNeb}, 
based on the measured $T_{\rm e}$([$\oiii$]) 
and either measured or inferred
$T_{\rm e}$([$\oii$]).
The resulting metallicities range 
from 12 + log(O/H) = 7.2 to 8.4,
indicating that these high-redshift
galaxies are relatively metal-poor,
in agreement with prior JWST findings.

\item We present, for the first time, 
the mass-metallicity relation (MZR) 
for galaxies in the redshift range 
$3 < z < 10$, with metallicities 
entirely derived using the 
direct-$T_e$ method.
To estimate the average MZR, 
we divided our sample into 
three stellar mass bins: 
log($M_{\ast}/M_{\odot}$) = 
7–8, 8–9, and 9–10,
ensuring that each bin contained 
at least 15 galaxies. 
Our analysis yielded a
best-fit slope of 
$\gamma = 0.211 \pm 0.120$ and
a metallicity intercept of
$Z_{10} = 7.986 \pm 0.205$. 
The slope and normalization of
our MZR are consistent within
1$\sigma$ uncertainty with previous 
high-redshift studies where
metallicities were derived 
using local calibrations, 
including JADES + Primal 
\citep{2024arXiv240807974S}, 
CEERS \citep{2023ApJS..269...33N}, 
and JADES \citep{2024A&A...684A..75C}.
However, our best-fit curve is
approximately 0.2 dex lower 
at $M_{\ast} = 10^8 M_{\odot}$ 
compared to these studies. 
Additionally, when compared with
MZR studies at lower redshifts, 
we find that metallicities for a
given stellar mass in our 
sample are noticeably lower 
than those reported at $z \sim 0$ 
by \citet{2013ApJ...765..140A} 
and at $z \sim 0.08$ by
\citet{2020MNRAS.491..944C}.
We estimate the scatter,
\(\sigma_{\text{scatter}}\), 
in our MZR to be approximately
0.09 dex,
which aligns well with the 
intrinsic scatter of 0.08 dex 
observed in low-stellar-mass 
(log \(M_\star / M_\odot \lesssim 9\)) 
galaxies at redshifts \(3 < z < 10\) 
\citep{2024A&A...684A..75C}.

\item We utilized our full sample 
of 67 galaxies (42 from this
study and 25 from previous literature)
within the redshift range 
$3 < z < 10$ to develop new 
empirical metallicity calibrators 
tailored to high-redshift galaxies.
A polynomial fitting approach, 
log($R$) = $\sum c_n x^n$, 
was used to derive calibrators
based on six line ratios: 
R3, R2, R23, Ne3O2, O32, and O3N2,
spanning metallicities of 
12 + log(O/H) = 7.2–8.4. 
The best-fit polynomials for
each line ratio are presented 
in Table \ref{tab:calibrators}.

\item
We tested the novel metallicity 
calibration relation proposed by 
\citet{2024A&A...681A..70L} 
for local galaxies, 
defined as $\hat{R}$ = 
cos($\Phi$) log(R2) + sin($\Phi$) log(R3) 
with $\Phi = 61.82^\circ$, 
on our high-redshift sample. 
For this relation, 
we found an RMS scatter 
of $\sigma_{\rm RMS} = 0.14$ when 
comparing the best-fit Rcap values 
from \citet{2024A&A...681A..70L}
with our observational data.
Following the methodology of 
\citet{2024A&A...681A..70L}, 
we optimized the $\Phi$ parameter
for our high-redshift sample,
finding an optimal $\Phi$ = 
$79.8 \pm 2.4^\circ$, which yields
$\hat{R}$ = 0.18 log(R2) + 0.98 log(R3). 
For our optimized Rcap, the RMS 
scatter reduced significantly
to $\sigma_{\rm RMS} = 0.06$, 
indicating a tighter calibration
compared to the original Rcap for
local galaxies and confirming a 
large intrinsic scatter among 
high-redshift galaxies.

\end{itemize}

\begin{acknowledgements}
We extend our heartfelt gratitude to Dr. Lisa Kewley for her invaluable insights and stimulating discussions.
PC acknowledges support from NASA Chandra grants GO3-24124X 16619325, GO3-24024X 16619349, and GO4-25094X 16619354.

This work is based on observations conducted with the NASA/ESA/CSA James Webb Space Telescope (JWST). The data were obtained from the Mikulski Archive for Space Telescopes (MAST) hosted by the Space Telescope Science Institute (STScI). STScI is operated by the Association of Universities for Research in Astronomy, Inc., under NASA contract NAS 5-03127 for JWST operations.  
We acknowledge the JADES repository \citep{2023ApJS..269...16R}. The JADES data products used in this study were retrieved from the Data Release 3 archive, available at \url{https://jades-survey.github.io/scientists/data.html}.  
Additionally, the Primal Survey data products were obtained from the DAWN JWST Archive (DJA). The DJA is an initiative of the Cosmic Dawn Center, funded by the Danish National Research Foundation under grant DNRF140.
\end{acknowledgements}



\bibliographystyle{mnras}
\bibliography{sample631} 

\appendix


\begin{table*}[h!!]
   \caption{GOODS-S and GOODS-N JWST/NIRSpec Sample of Galaxies
   Analysed in this Paper
   \citep{2024arXiv240406531D}.}
    \centering
    \setlength{\tabcolsep}{5pt}
    \begin{tabular}{ccccccccc}
    NIRSpec ID & RA & Dec & $z$ & log($M_{\ast}/M_{\odot}$) &  12+log(O/H) (direct) & T$_{\rm e}$ (K)\tablenotemark{$\dagger$}\\
\hline
\hline
54612 & 53.14468 & -27.771185 & 3.08 & 9.05 $\pm$ 0.05 & 8.18 $\pm$ 0.15 & 12441 $\pm$ 1211 \\
39898 & 53.1324005 & -27.8091354 & 3.17 & 9.68 $\pm$ 0.03 & 7.47 $\pm$ 0.17 & 23929 $\pm$ 2177 \\
26113 & 189.0996212 & 62.2635629 & 3.23 & 8.56 $\pm$ 0.07 & 7.26 $\pm$ 0.1 & 21483 $\pm$ 2005 \\
4550 & 189.1924782 & 62.2388249 & 3.24 & 8.15 $\pm$ 0.06 & 7.68 $\pm$ 0.11 & 18735 $\pm$ 2103 \\
56785 & 189.3147109 & 62.2023856 & 3.44 & 9.25 $\pm$ 0.07 & 7.57 $\pm$ 0.09 & 23800 $\pm$ 2082 \\
1137 & 189.105766 & 62.283372 & 3.66 & 9.07 $\pm$ 0.09 & 7.74 $\pm$ 0.09 & 18503 $\pm$ 1545 \\
3683 & 189.0936083 & 62.2918914 & 3.66 & 8.31 $\pm$ 0.04 & 7.79 $\pm$ 0.15 & 15962 $\pm$ 1725 \\
10009453 & 53.1787077 & -27.7989084 & 3.71 & 7.56 $\pm$ 0.02 & 7.52 $\pm$ 0.13 & 19818 $\pm$ 2591 \\
1048 & 189.0583155 & 62.2725583 & 3.87 & 7.98 $\pm$ 0.04 & 7.40 $\pm$ 0.06 & 21767 $\pm$ 1100 \\
15529 & 189.215044 & 62.2770075 & 3.87 & 8.75 $\pm$ 0.03 & 8.29 $\pm$ 0.06 & 12041 $\pm$ 521 \\
902 & 189.1932763 & 62.2537271 & 4.06 & 8.50 $\pm$ 0.02 & 7.99 $\pm$ 0.12 & 14159 $\pm$ 1238 \\
1165 & 189.111861 & 62.2877299 & 4.38 & 8.30 $\pm$ 0.02 & 8.20 $\pm$ 0.14 & 12162 $\pm$ 1170 \\
16553 & 189.1436028 & 62.2805455 & 4.38 & 8.46 $\pm$ 0.02 & 7.78 $\pm$ 0.12 & 16424 $\pm$ 1512 \\
11836 & 189.2205875 & 62.2636751 & 4.41 & 8.29 $\pm$ 0.02 & 7.59 $\pm$ 0.05 & 18470 $\pm$ 975 \\
10000865 & 189.2629857 & 62.2501062 & 4.41 & 8.56 $\pm$ 0.02 & 7.75 $\pm$ 0.11 & 17023 $\pm$ 1738 \\
206035 & 53.1581721 & -27.7864763 & 4.77 & 8.30 $\pm$ 0.02 & 7.78 $\pm$ 0.16 & 16672 $\pm$ 2248 \\
61321 & 53.154274 & -27.7524204 & 4.84 & 8.31 $\pm$ 0.02 & 7.92 $\pm$ 0.06 & 17116 $\pm$ 912 \\
10009642 & 53.1926798 & -27.7842212 & 4.84 & 8.86 $\pm$ 0.02 & 8.02 $\pm$ 0.05 & 16122 $\pm$ 700 \\
920 & 189.0981446 & 62.2555524 & 4.88 & 8.65 $\pm$ 0.03 & 7.95 $\pm$ 0.18 & 14910 $\pm$ 2109 \\
70920 & 189.2509624 & 62.1603815 & 5.04 & 8.29 $\pm$ 0.02 & 7.43 $\pm$ 0.04 & 23973 $\pm$ 1119 \\
607 & 189.1169473 & 62.2220788 & 5.18 & 8.31 $\pm$ 0.03 & 7.59 $\pm$ 0.15 & 21021 $\pm$ 2046 \\
721 & 189.1153168 & 62.2340987 & 5.18 & 7.58 $\pm$ 0.015 & 7.45 $\pm$ 0.18 & 20728 $\pm$ 2267 \\
59412 & 189.1563218 & 62.2100022 & 5.18 & 8.81 $\pm$ 0.02 & 7.67 $\pm$ 0.19 & 18812 $\pm$ 3021 \\
79349 & 189.2096823 & 62.207252 & 5.18 & 7.83 $\pm$ 0.03 & 7.51 $\pm$ 0.12 & 21021 $\pm$ 2491 \\
131737 & 53.1990421 & -27.7725801 & 5.89 & 7.97 $\pm$ 0.05 & 7.47 $\pm$ 0.06 & 21050 $\pm$ 1102 \\
78891 & 189.2258239 & 62.2042147 & 6.55 & 8.68 $\pm$ 0.03 & 8.02 $\pm$ 0.07 & 14569 $\pm$ 821 \\
1967 & 189.1650306 & 62.3001933 & 6.56 & 8.49 $\pm$ 0.02 & 7.50 $\pm$ 0.17 & 21347 $\pm$ 2140 \\
38428 & 189.1792747 & 62.2758955 & 6.71 & 9.47 $\pm$ 0.02 & 7.47 $\pm$ 0.06 & 20522 $\pm$ 1100 \\
38432 & 189.18617 & 62.2708636 & 6.72 & 8.87 $\pm$ 0.03 & 7.88 $\pm$ 0.1 & 16107 $\pm$ 1415 \\
38420 & 189.17514 & 62.2822634 & 6.73 & 8.28 $\pm$ 0.02 & 7.51 $\pm$ 0.15 & 19854 $\pm$ 2839 \\
18536 & 189.1553143 & 62.2864715 & 6.81 & 8.21 $\pm$ 0.02 & 7.80 $\pm$ 0.1 & 16892 $\pm$ 1495 \\
20213084 & 53.1589064 & -27.765076 & 8.49 & 8.09 $\pm$ 0.03 & 7.41 $\pm$ 0.02 & 20761 $\pm$ 512 \\
265801\tablenotemark{$\ast$} & 53.1124351 & -27.7746258 & 9.43 & 8.17 $\pm$ 0.04 & 7.36 $\pm$ 0.08 & 21176 $\pm$ 1916 \\

\hline
\vspace*{-\baselineskip}
    \end{tabular}
    \label{tab:goods}
    \tablenotetext{\ast} { Metallicity of JADES-GS-z9-0 at z = 9.4327 is also reported by \citet{2024arXiv240702575C},
    consistent with our new measurements.}
    \tablenotetext{\dagger}{ Temperature measured using [$\oiii$]$\lambda$4363 emission line.}
\end{table*}

\begin{table}[h!!]
   \caption{PRIMAL JWST/NIRSpec Sample of Galaxies
   Analysed in this Paper
   \citep{2024arXiv240402211H}.}
    \centering
    \setlength{\tabcolsep}{5pt}
    \begin{tabular}{ccccccccc}
    NIRSpec ID & RA & Dec & $z$ & log($M_{\ast}/M_{\odot}$) &  12+log(O/H) (direct) & T$_{\rm e}$ (K)\tablenotemark{$\dagger$}\\
\hline
\hline
4385 & 215.1795990 & 53.0620667 & 3.42 & 8.20 $\pm$ 0.04 & 8.08 $\pm$ 0.12 & 14596 $\pm$ 1339 \\
13491 & 53.1501095 & -27.8197035 & 3.47 & 8.37 $\pm$ 0.02 & 7.69 $\pm$ 0.03 & 14815 $\pm$ 1166 \\
3585 & 215.0232093 & 53.0079722 & 3.87 & 8.59 $\pm$ 0.05 & 7.50 $\pm$ 0.08 & 23383 $\pm$ 1949 \\
9489 & 53.1687170 & -27.8151743 & 4.02 & 7.99 $\pm$ 0.02 & 7.90 $\pm$ 0.09 & 14346 $\pm$ 1062 \\
40066 & 3.5997164 & -30.4318948 & 4.02 & 9.40 $\pm$ 0.10 & 7.79 $\pm$ 0.04 & 14536 $\pm$ 316 \\
1173 & 215.1542076 & 52.9558470 & 5.00 & 9.38 $\pm$ 1.12 & 7.64 $\pm$ 0.10 & 18642 $\pm$ 1870 \\
110000 & 3.5706428 & -30.4146380 & 5.76 & 9.22 $\pm$ 0.27 & 7.55 $\pm$ 0.08 & 17397 $\pm$ 1243 \\
9842 & 53.1540870 & -27.7660620 & 5.80 & 7.62 $\pm$ 0.03 & 7.31 $\pm$ 0.15 & 22186 $\pm$ 2881 \\
2782 & 214.8234525 & 52.8302813 & 5.24 & 8.53 $\pm$ 0.05 & 7.28 $\pm$ 0.04 & 22777 $\pm$ 900 \\
689 & 214.9990525 & 52.9419767 & 7.55 & 8.70 $\pm$ 0.01 & 7.62 $\pm$ 0.11 & 19339 $\pm$ 1700 \\

\hline
    \end{tabular}
    \label{tab:primal}
    \tablenotetext{\dagger}{ Temperature measured using [$\oiii$]$\lambda$4363 emission line.}
\end{table}

\end{document}

%% file: macros.tex
\newcommand{\Cloudy}{\textsc{Cloudy}}

\font\manual=manfnt at 7pt \def\dbend{\hbox{\raise0.9ex\hbox{\manual\char127\hspace{0.6em}}}}

\providecommand{\e}[1]{\ensuremath{\times 10^{#1}}}

\newcounter{INTERNALionstage}

\def\gtsim{\mathrel{\hbox{\rlap{\hbox{\lower4pt\hbox{$\sim$}}}\hbox{$>$}}}}
\def\lesssim{\mathrel{\hbox{\rlap{\hbox{\lower4pt\hbox{$\sim$}}}\hbox{$<$}}}}

\def\A{{\rm\thinspace \AA}}

\def\g{{\rm\thinspace g}}

\def\K{{\rm\thinspace K}}

\def\m{{\rm\thinspace m}}

\def\s{{\rm\thinspace s}}

\def\W{{\rm\thinspace W}}

\def\hii{\mbox{{\rm H~{\sc ii}}}}

\def\heii{\mbox{{\rm He~{\sc ii}}}}

\def\nii{\mbox{{\rm N~{\sc ii}}}}

\def\oii{\mbox{{\rm O~{\sc ii}}}}
\def\oiii{\mbox{{\rm O~{\sc iii}}}}
\def\oiv{\mbox{{\rm O~{\sc iv}}}}

\def\neiii{\mbox{{\rm Ne~{\sc iii}}}}

\def\sii{\mbox{{\rm S~{\sc ii}}}}

\def\h0{\mbox{{\rm H}$^0$}}
\DeclareMathAlphabet{\vib}{OML}{cmm}{m}{it}

